\documentclass[12pt]{article}
\usepackage{amsmath}
\usepackage{amssymb}
\usepackage{amsthm}
\usepackage{cite}
\usepackage{comment}
\usepackage{graphicx}
\usepackage{grffile}
\graphicspath{{figures/}}
\usepackage{hyperref}
\usepackage{verbatim}
\usepackage{color}
\usepackage{float}
\usepackage{caption}
\usepackage{subcaption}
\usepackage[section]{placeins}
\edef\restoreparindent{\parindent=\the\parindent\relax}
\usepackage{parskip}
\restoreparindent
\usepackage[normalem]{ulem}
\usepackage{fullpage}
\usepackage[sort&compress,numbers]{natbib}

\newcommand{\be}{\begin{equation}}
\newcommand{\ee}{\end{equation}}
\newcommand{\bea}{\begin{eqnarray}}
 \newcommand{\eea}{\end{eqnarray}}
 
 \newcommand{\re}{\mathbb R}
 \newcommand{\bbC}{\mathbb C}
 \newcommand{\Zon}{Z_{\Omega_n}}
 \newcommand{\twod}{\mathrm{2 d}}
 \newcommand{\dimm}{\mathbf{d}}
 \newcommand{\dor}{{d}}
\newcommand{\mink}{\mathbb M}
\newcommand{\mL}{\mathcal L}

\newcommand{\ocylmn}{\Omega^{(m,n)}}

\newcommand{\mlcylm}{\mL^{(m)}}
\newcommand{\cEcylmn}{\mathcal{E}^{(m,n)}}
\newcommand{\bE}{\bar E}
\newcommand{\bbe}{\bar e} 
\newcommand{\bt}{\bar t}
\newcommand{\bx}{\bar x}
\newcommand{\bds}{S_{\mathrm{BD}}}
\newcommand{\cZ}{\mathcal Z}
\newcommand{\av}[1]{\langle {#1} \rangle}
\newcommand{\vol}{\mathrm{vol}}
\newcommand{\ma}{M_a}
\newcommand{\ga}{\eta_a} 
\newcommand{\mb}{M_b}
\newcommand{\gb}{\eta_b}
\newcommand{\mab}{M_{a,b}}
\newcommand{\gab}{\eta_{a,b}} 
\newcommand{\bcd}{\beta_c^{(\dimm)}}
\newcommand{\bctwo}{\beta_c^{(2)}}

\begin{document}
\title{Dimensionally Restricted  Causal Set Quantum Gravity: \\ Examples  in  Two and Three  Dimensions }
\author{William J.\ Cunningham${}^1 $ and Sumati Surya${}^2$\\ ${}^1${\it{\small{Perimeter Institute for Theoretical Physics, 31 Caroline St. N., Waterloo, Ontario, Canada }}}
 \\ ${}^2${\it{\small{Raman Research Institute, CV
    Raman Ave., Sadashivanagar, Bangalore, 560080, India}}}}
\date{}
\maketitle
\begin{abstract}
  We study dimensionally restricted non-perturbative causal set quantum dynamics in $2$ and  $3$ spacetime
  dimensions with non-trivial global spatial topology. The causal set sample space is generated from causal
  embeddings into  spacetime lattices with  global spatial topology $S^1$ and $T^2$ in $2$ and $3$  
  dimensions, respectively.  The quantum gravity partition function over these  sample spaces is studied using Markov
  Chain Monte Carlo (MCMC) simulations after  analytic
  continuation.   In both  $2$ and  $3$ dimensions we find a
  phase transition that separates the dominance of the action from that of  the entropy. The action dominated phase is
  characterised  by ``layered''  posets with a high degree of connectivity,  while the causal sets in the entropy
  dominated phase are manifold-like. This phase transition is similar in character to that seen for the sample space of $2$-orders,
  which are topologically trivial, hence suggesting that this is a generic feature of dimensionally restricted sample
  spaces. The  simulations use a newly developed framework for causal set MCMC calculations.  Ours is 
  the first implementation of  a causal set dynamics restricted to  $3$ dimensions.


\end{abstract}
\clearpage

\section{Introduction}

Causal set theory (CST) makes the bold assumption that the spacetime continuum is an approximation of an underlying locally finite partially
ordered set \cite{blms, valdivia,fayreview, joereviewone,joereviewtwo, walldenreview,livingreviews}. While the continuum
approximation of the theory is deeply rooted in Lorentzian geometry~\cite{hkm,malament},  there are fewer known
constraints on the dynamics. In the most conservative approach, the partition function  over 
Lorentzian geometries is replaced by the weighted sum over the  \emph{sample space}\footnote{Refer to Appendix
  \ref{app:defs} for  a list of standard definitions in CST.} $\Omega_n$ of $n$-element causal sets 
\begin{equation}
  \Zon \equiv  \sum_{c\in \Omega_n} \mu(c) \,,
  \label{eq:partn} 
\end{equation} 
where $\mu(c) \in \bbC$ is the measure. Fixing $n$ is analogous to fixing the total spacetime volume in the spirit of unimodular
 gravity \cite{unruhwald,lambdatwo}, and is a natural choice for causal sets. In analogy with quantum field theory, one moreover expects that the measure takes
the specific form
\begin{equation}
   \label{eq:bdmeasure} 
   \mu(c) = \exp(i \bds^{(\dimm)}(c)/\hbar) \,,
\end{equation}
where $S^{(\dimm)}_{BD}(c)$ is the $\dimm$-dimensional discrete Einstein-Hilbert or \emph{Benincasa-Dowker (BD)
action}~\cite{bd,dg,glaser}. 
 
  Even in this most conservative setting, the dynamics encoded in $\Zon$ is considerably more inclusive than in 
  other discrete approaches to quantum  gravity . For one, $\Omega_n$ is entropically dominated by  causal sets that are nothing like
  any spacetime. These \emph{Kleitman-Rothschild} (KR) posets have  three
  \emph{layers} or moments of ``time'' and their number grows at least like $\sim
  2^{n^2/4} $ \cite{kr}.  There is moreover  a  hierarchy
  of sub-dominant posets which are also not manifold-like  \cite{dharone,dhartwo}.  The action must therefore be
  able to counter this  entropy in the
  classical limit of the theory, where one expects the continuum to be emergent.  $\Omega_n$ also  contains causal sets
  that are approximated by spacetime geometries of  {\it any}  dimension,  for large enough $n$. Thus, the action must
  additionally ensure that  $\dimm=4$
  dominates  in the classical limit of the theory.

  The choice of measure in the partition function $\Zon$ is therefore very important.  Although the measure Eqn.~\ref{eq:bdmeasure} appears natural from the perspective of local quantum
  field theory,  there is no fundamental or order theoretic motivation for this particular choice.  Since the
  continuum is replaced by the discretum of causal sets, with the space of causal sets including those that are not
  continuum-like,  one expects the Einstein-Hilbert action to be an effective rather than a fundamental 
  action (see for example \cite{astridcg}).  Thus the measure could include more fundamental order theoretic contributions
  that cannot be guessed at from the continuum. An example of this is the weight coming from the  number of possible 
  relabellings of the causal set. In \cite{carliploomis}  such a counting measure was used to  help suppress 
  the  contribution of  the  class of (non-continuum-like) bilayer posets to the partition function. 

  In particular, in the continuum the formal path integral of quantum gravity  is defined over geometries of a fixed dimension $\dimm$ with
  the measure given in terms of  the Einstein-Hilbert action of the same dimension. The  choice of
  the continuum-inspired measure,  Eqn.~(\ref{eq:bdmeasure}),   therefore requires a choice of $\dimm$ from the outset. In
  order to mimic the formal continuum path integral, $\Omega_n$ should be further restricted   
  to the set of $n$-element  causal sets whose continuum limit, if it exists, is a $\dimm$ dimensional spacetime.

  Even if it were  possible to do this dimensional ``cherry picking'',  the naive choice of measure,  Eqn.~(\ref{eq:bdmeasure}),
  ignores order-theoretic factors that could arise from causal set theory.  While the measure must satisfy discrete covariance or
  \emph{order invariance}, the continuum can only act as a rough guide.    
  For example, it is  possible that due to the nature of the continuum approximation in
  CST (see \cite{blms, valdivia,fayreview, joereviewone,joereviewtwo, walldenreview,livingreviews}),  there are  
  relative weights  between different continuum-like
  causal sets which cannot be obtained purely from a continuum theory.  Also, since the sample space must contain
  causal sets that are not manifold-like, there can also be a  relative
 weight between  manifold-like and non-manifold-like causal sets.  Importantly, hidden in the measure of
 Eqn.~(\ref{eq:partn}) 
  is the  choice of sample space, which may come   with its own preferred choice of measure.   Thus, in trying to
  mimic a continuum path integral in CST, a myriad choices can be made in principle.  

  Our aim in this work is to generate dimensional and topological restrictions $\Omega_n(M,g)$ of $\Omega_n$ associated with
  a continuum spacetime $(M,g)$  in order to study dimensionally restricted causal set dynamics.  For a causal diamond $D_2$ in $\dimm=2$ Minkowski spacetime $\mink^2$,
 it was shown that  the  continuum dimension coincides with a natural order-theoretic dimension 
 defined for \emph{$\dor$-orders} \cite{2dorders,2dqg}.  These  $2$-orders (or $\twod$-orders)  admit a simple representation
in  the light-cone coordinates, with the elements occupying sites in a 
  regular light-cone lattice in  $(D_2,\eta)$.   While the sample space of $n$-element $2$-orders does contain  those that are not manifold-like, it is dominated by
  causal sets that  are   approximated by the flat causal diamond $(D_2,\eta)$ \cite{winkler,es} with trivial
  global spatial 
topology.  Entropically, however, it is a much easier beast to handle than $\Omega_n$ and hence it has been
possible to do an extensive study of the 
  continuum-inspired dynamics for $2$-orders \cite{2dqg,hh,fss}.

 In higher dimensions,  the $\dor$-orders do not 
correspond to causal sets that  are continuum-like, since the associated ``light-cones''  for a
$\dor$-order are  in fact hypercubes, with cross-sections given by $(\dimm-1)$-simplices, rather than $(\dimm-1)$ spheres. Hence restricting to the sample space of
  $d$-orders would take us further away from our goal of mimicking the continuum path integral. 

  In order to make progress in finding a dimensional restriction for $\dimm >2$ and also to explore non-trivial global topology
  in $\dimm\geq 2$, it is helpful to view the $2$-orders as causal sets obtained from \emph{causal embeddings}  into the
  causal diamond $(D_2,\eta)$. Let $\Omega_n(D_2,\eta)$ denote the sample space of causal sets that causally embed into
  $(D_2,\eta)$.  To generate $\Omega_n(D_2,\eta)$ on the computer, consider a regular light-cone lattice on
  $D_2$ with $m$ sites.  Starting from an $n$-site filling of this lattice and using the induced causal relations between the sites
  in $(D_2,\eta)$, we obtain an $n$-element  causal set which causally embeds into $(D_2,\eta)$. In the limit of large
  $m$, the set of causal sets obtained this way $\sim \Omega_n(D_2,\eta)$. In the specific case, with   $m=n^2$,  and such
  that no two filled sites have the same light-cone coordinates  (i.e.,  they are not null related)
  we recover the sample space of $2$-orders, which can therefore be viewed as an approximation of  $\Omega_n(D_2,\eta)$.

  In this work we propose such a lattice-inspired model for generating dimensionally and topologically restricted sample
  spaces $\Omega_n(M,g)$ from a topologically non-trivial but finite volume (region of a ) spacetime $(M,g)$.  In
  Sec.~\ref{sec:sample} we describe how to obtain an approximation $\Omega^{(m,n)}$ of  the sample space on the
  computer  via a latticisation  of $(M,g)$ into $m$-sites. We focus 
  on two specific
  examples: (a) $(\ma,\ga)$ is the flat cylinder spacetime in $\dimm=2$, with global spatial topology $S^1$ and (b) $(\mb,\gb)$
  is the flat toroidal
  spacetime in $\dimm=3$, with global spatial topology $T^2$.  In order to generate $\Omega^{(m,n)}$ we use a lattice-gas
  inspired Markov Chain
  Monte Carlo (MCMC) 
  algorithm to obtain the expectation values of various observables, or \emph{order-invariants}.  
  We demonstrate that  for large enough $m$,  $\Omega_{a,b}^{(m,n)} \sim \Omega_n(\mab, \gab)$   
 by noting that  the expectation values of various observables calculated in the MCMC simulations converge to the same
 values.  In Sec.~\ref{sec:dynamics} we set up a continuum-inspired dynamics on  $\Omega_n(\mab,\gab)$, with any extra  
order theoretic measure  coming  entirely from the choice of the sample space.  The introduction of a parameter $\beta$
  analogous to an  inverse temperature admits an analytic continuation,  so that the lattice-gas MCMC 
  algorithms may be used to simulate the non-perturbative dynamics. 

  Section \ref{sec:2d3d} contains the results of the lattice-gas MCMC simulations for $(\mab,\gab)$. 
We find a phase transition which 
  exhibits  the tussle between entropy and action in
  the  Lorentzian statistical geometry setting\footnote{In \cite{carliploomis}  this has been studied for the quantum
    path integral  for the specific case of bilayer posets.} which echoes that found for $2$-orders  \cite{2dqg}. In
  Section~\ref{sec:conclusions} we discuss some of the open questions and ongoing related work. This study sets the
  stage for a new class of investigations into causal 
  set dynamics, which can bridge the gap between the deep quantum,  order-theoretic regime and the framework of the continuum 
  path integral.   Finally, Appendix ~\ref{app:defs} contains some of the standard definitions used throughout
  this work, italicised when they first appear in the text for ease of reading, and  Appendix~\ref{app:of} contains a
  calculation of the ordering fraction for  causal set approximated by  the cylinder spacetime.

  \section{\texorpdfstring{Generating new sample spaces $\Omega_n(M,g)$}{Generating new sample spaces}}
  \label{sec:sample}

As mentioned in the introduction, the
  sample space $\Omega_n(M,g)$ of causal sets  that  causally embed into a finite volume (preferably globally hyperbolic) region
  $(M,g)$  of a $\dimm$-dimensional  spacetime  includes those that are non-manifold-like as well as those whose
  continuum approximation differs significantly from $(M,g)$. The only point of similarity for the 
  manifold-like causal sets in $\Omega_n(M,g)$  is their global topology.  A quantum dynamics on $\Omega_n(M,g)$ would
  thus be  an example of  a $\dimm$-dimensional causal set quantum gravity with global topology $M$.  Our goal in this
  section is to see how to generate $\Omega_n(M,g)$ on the computer,  using an
  appropriate latticisation of  $(M,g)$.   

\begin{figure}[!t]
\centering
\includegraphics[width=0.7\textwidth]{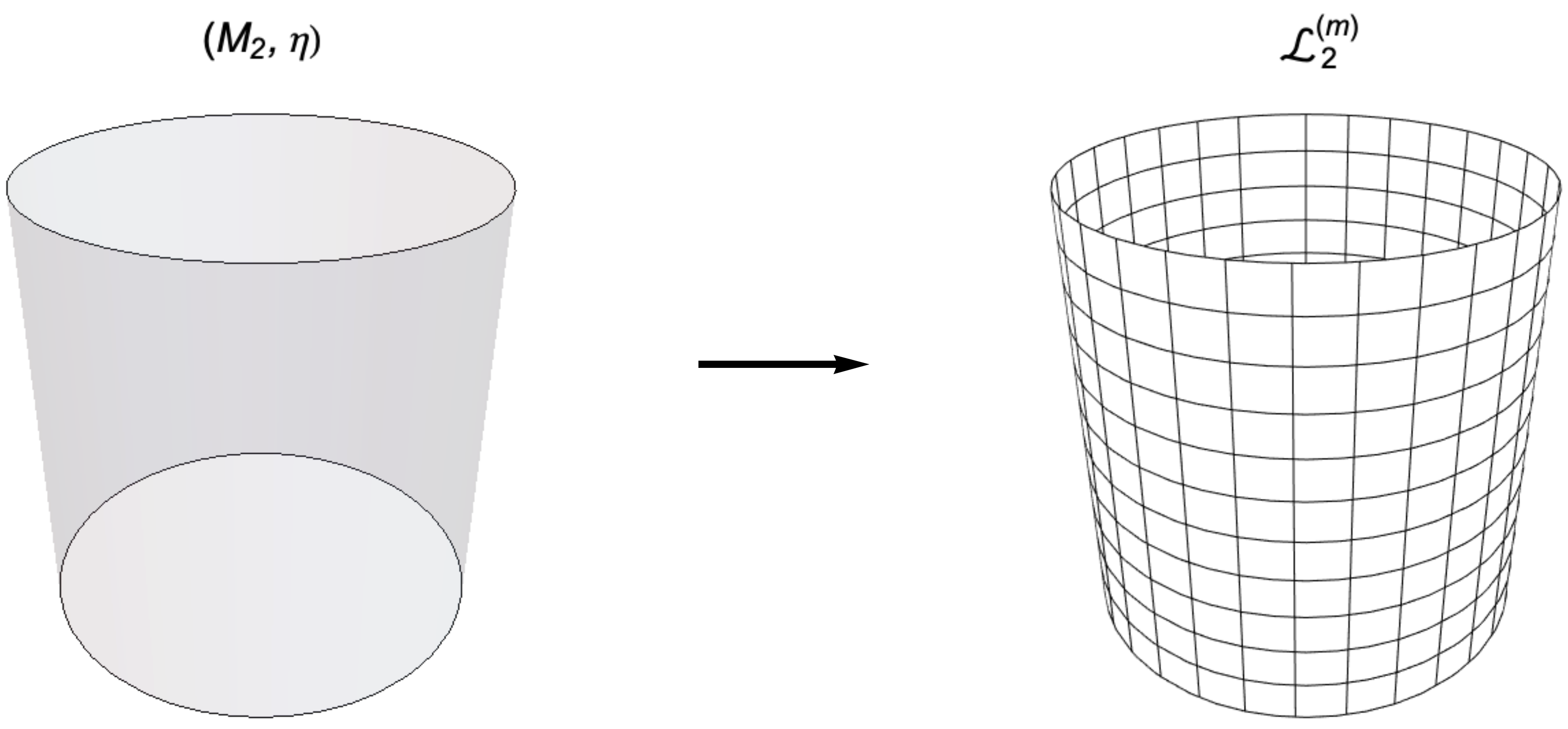}
\caption{The latticisation $\mathcal{L}^{(m)}$ of the spacetime $(\ma, \eta)$ into a grid of height $h$ and
  width $w$, where $m=h\times w$.}
\label{fig:lattice}
\end{figure}
 
  \begin{figure}[!t]
\centering
\includegraphics[width=0.3\textwidth]{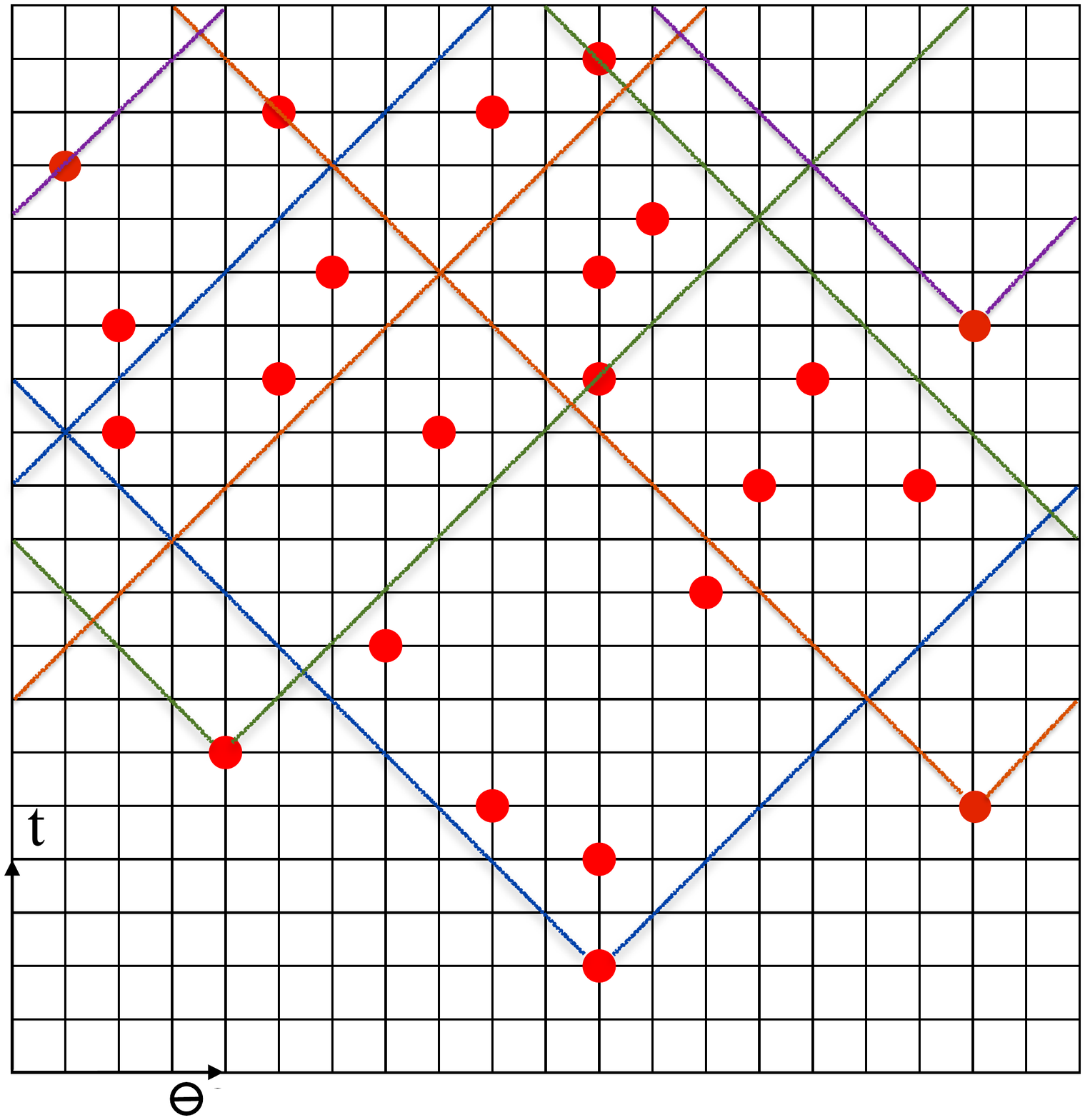}\hskip 0.5cm\includegraphics[width=0.3\textwidth]{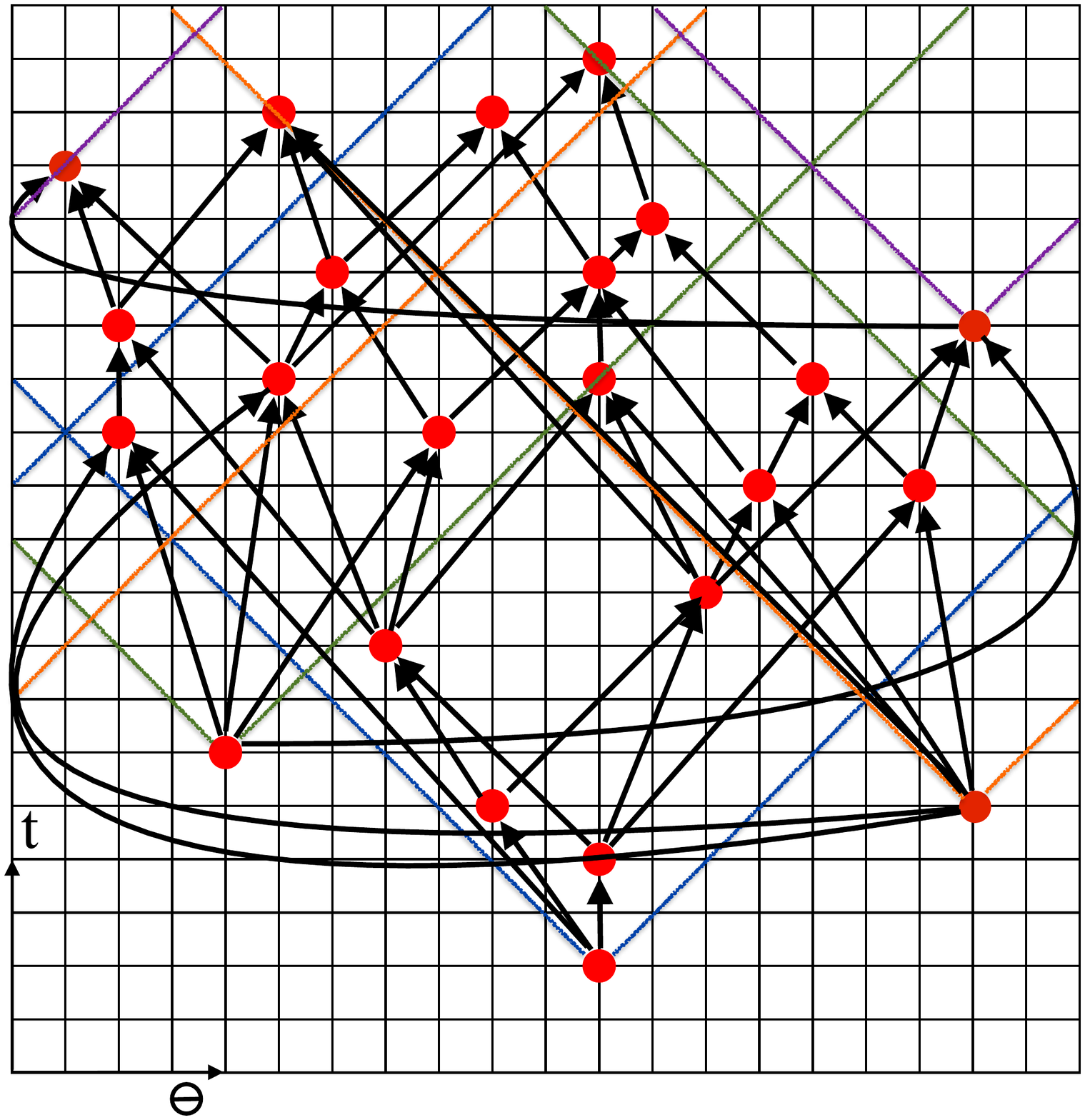}\hskip 0.5cm\includegraphics[width=0.3\textwidth]{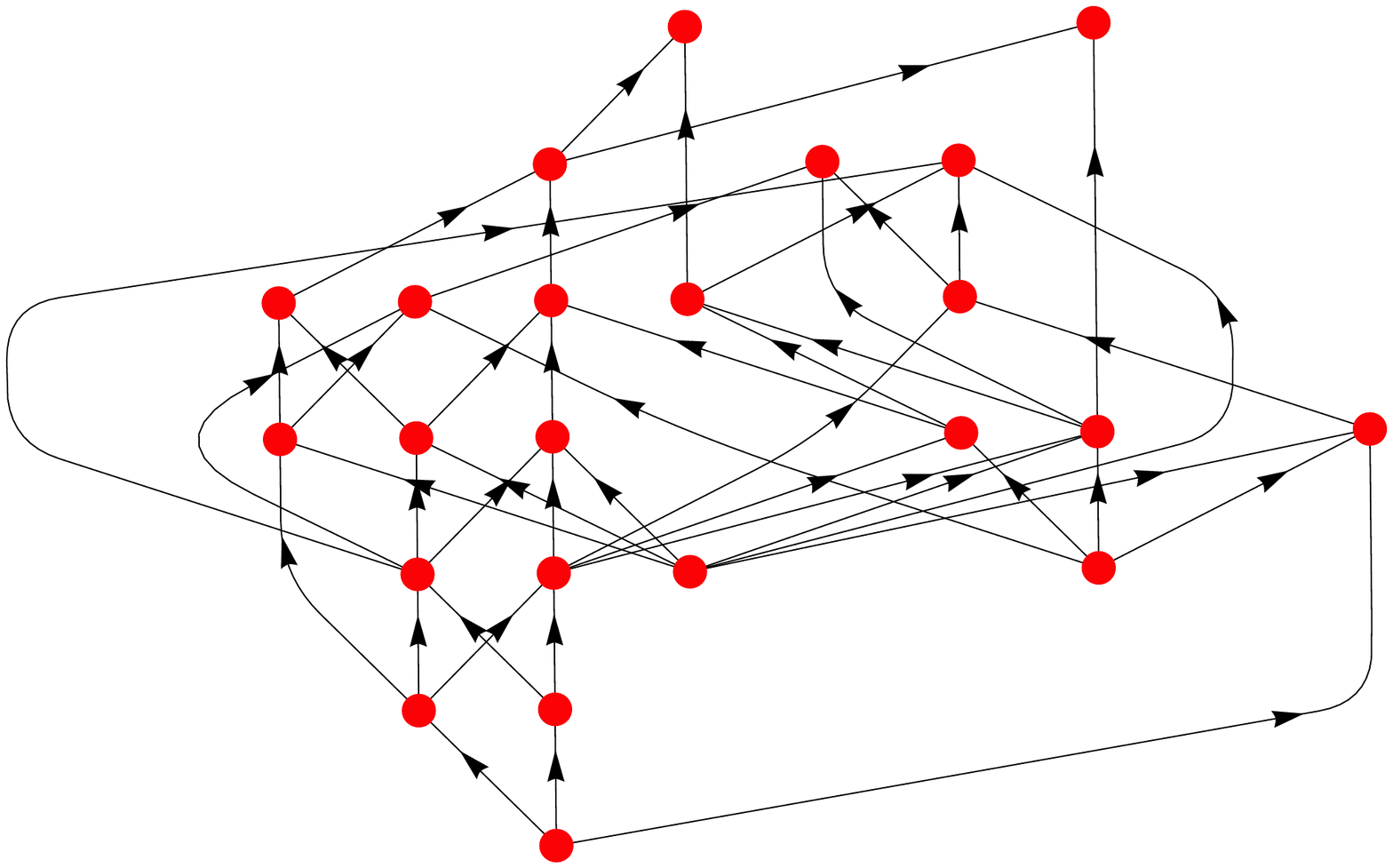}
\caption{(i) An $n$-site filling of an $m$-site cylindrical lattice (the spatial edges are identified). The light-cones for a
  few of the elements are shown and can be seen to  wrap around the cylinder. (ii) The links or irreducible relations between elements are obtained from the causal
  structure of the lattice.  The curved lines denote the relations obtained by going ``around''  the cylinder, due to
  the 
  topological non-triviality of the causal structure. (iii) The resulting  Hasse diagram for the  causal set.}

\label{fig:latticecausalsets}
\end{figure}
 To make the discussion concrete, we consider the  flat spacetimes $(M_\dimm,\eta)$ with toroidal global spatial topology 
\begin{eqnarray} 
   M_\dimm &\sim&  I \times T^{\dimm-1}=\re \times \underbrace{S^1 \times ..\ldots   S^1 }_{\dimm-1 }\,, \nonumber \\
  \eta_{\mu \nu}dx^\mu dx^\nu  &=& -dt^2+ \sum_{i=1}^{\dimm-1}d\theta_i^2, \quad \theta_i(\mathrm{mod})2\pi\,,\, t \in [0,h].
             \label{eq:cylspt} 
  \end{eqnarray} 
  These spacetimes admit a natural latticisation $\mlcylm_\dimm$ into an $h\times w^{\dimm-1} =m$ site cylindrical
  lattice, with each lattice site having the coordinates
  $(t,\theta_1, \ldots, \theta_{\dimm-1})$, where $t \in \{ 0, 1, \ldots , h\}$ and $ \theta_i = 2\pi k_i/w$,
  $k_i\in \{0,1, \ldots, w-1\}$, with the identification $\theta_i \sim \theta_i + 2 \pi$ (see   Fig.~\ref{fig:lattice}).  The lattice is parametrized by $m$ and also the \emph{aspect ratio}
  $\alpha=h/w$. Given the metric Eqn \ref{eq:cylspt}, the local light cones are therefore at $45^\circ$. 

  For any $n$-site filling of $\mlcylm_\dimm$,  one can construct a causal set using the causal
  relations in $(M_\dimm,\eta)$ induced on the $n$-sites as  illustrated in Fig.~\ref{fig:latticecausalsets}.  The
  causal future and past of every element is given via the metric Eqn \ref{eq:cylspt}. In
  Fig.~\ref{fig:latticecausalsets} there are three minimal elements which have no elements to their past  and three
  maximal elements with no elements to their future.

  In order to accommodate both the
  $n$-element \emph{chain} and the $n$-element \emph{antichain} in the sample space, we let $n\leq w,h$ so that $m \geq n^2$.  As discussed
  in the introduction, for the $2$-orders, $m=n^2$ and the global topology is $\sim D_2$.  Since we would like to capture the
  non-trivial global topology,  we also require that $h >w $, to allow at least a few null rays to wrap
  around the cylinder once. Ideally, $h$ should be significantly larger than $w$ for the full global topology to manifest
  itself in the sample space. 

\begin{figure}[!t]
\centering
\includegraphics[width=0.7\textwidth]{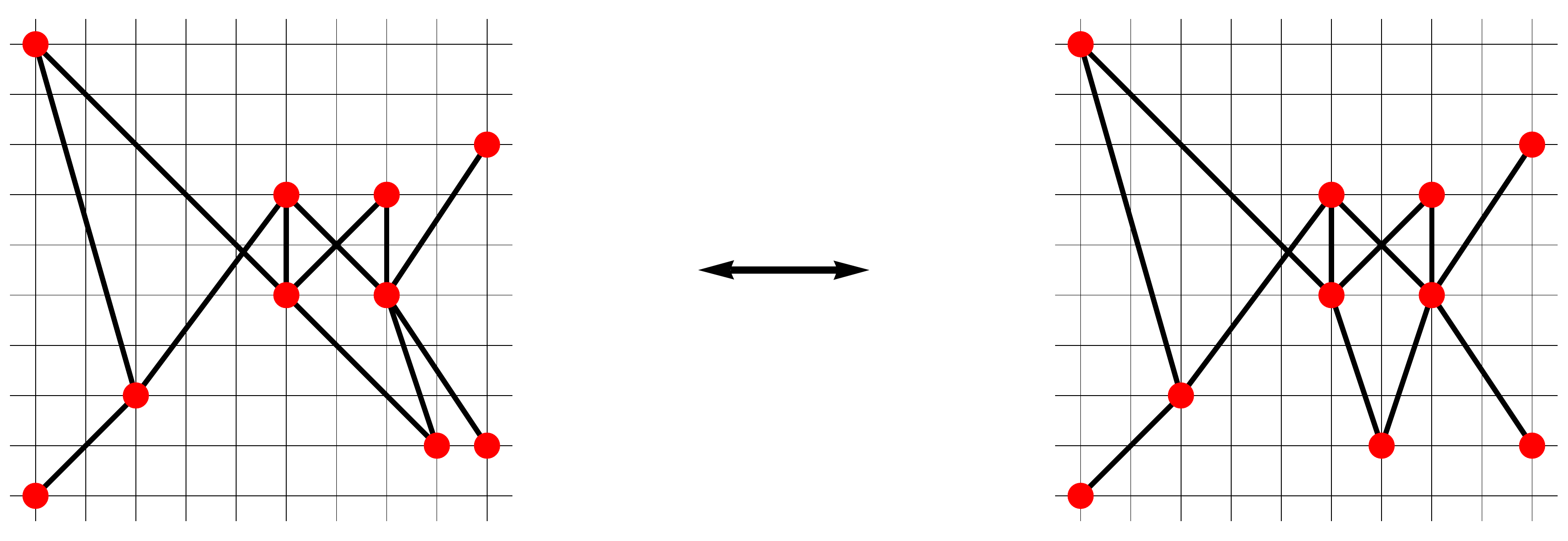}
\caption{An example of two distinct $n$-site fillings of an $m$-lattice which produce the same causal set. 
}
\label{fig:same_causet}
\end{figure}
  Let $\cEcylmn_\dimm$ be the set of all possible $n$-site fillings into $\mlcylm_\dimm$, so that  $|\cEcylmn_\dimm|=\binom{m}{n}$. Let 
  $\ocylmn_\dimm$ denote   the  set of  unlabelled causal sets, i.e.,  the \emph{(m,n)-orders}  obtained from $\cEcylmn_\dimm$.  For $n\ll m$, the map  $\cEcylmn_\dimm \rightarrow \ocylmn_\dimm$ is strictly non-injective since a small change in the location of a filled site in $\mlcylm_\dimm$ yields the same causal set in
  $\ocylmn_\dimm$ (Fig.~\ref{fig:same_causet}). Each causal set in $\ocylmn_\dimm$  is therefore typically  represented  multiple times in
  $\cEcylmn_\dimm$. For example, an $n$-element chain can be represented in multiple ways in $\cEcylmn_\dimm$ as can an
  $n$-element antichain. However, the number of possible representations for each  $c\in \ocylmn_\dimm$ is
  different. For a causal set obtained from a random selection of elements in $\cEcylmn_\dimm$ the redundancy will be 
  considerably smaller, since each element can occupy a smaller number of allowed sites without changing the causal
  relations. Thus, the map induces a non-uniform measure on $\ocylmn_\dimm$.  

  In general this measure depends not only on $n$, but also on $m$.  As $m$ increases,  however, the
  relative measures between the causal sets in $\ocylmn$ will become independent of $m$.  Roughly, this is because the number of empty sites surrounding 
 each filled site increases as the grid gets finer, but a move from a filled site to any one of these empty sites leaves the causal set
 unchanged. Thus the ``degeneracy'' of the representation of each  $c \in \ocylmn$  in $\cEcylmn$ should increase fairly
 uniformly with $m$.  We give evidence for this below in Fig.~\ref{fig:2d_variable_volume_zerob}  and also in Sec.~\ref{sec:2d3d}. 

 In particular, in  the limit of $m \rightarrow \infty$, with the lattice grid size going to zero,  we expect that
 $\ocylmn_\dimm \rightarrow  \Omega_n(M_\dimm,\eta)$.  Generating  $\cEcylmn_\dimm$ on the computer is therefore  one
 way to obtain a reasonable approximation of $\Omega_n(M_\dimm,\eta)$, as long as $m$ is sufficiently large. 
  
 One can  generate $\cEcylmn_\dimm$ by ``walking'' through every possible $n$-site filling of $\mlcylm_\dimm$, but of course this is
 computationally expensive,  since the number of possible fillings grows as $\sim m^n$.  It is therefore useful to devise
 a Markov Chain move that will effectively do the same job.  The move we describe is very simple and is inspired by
 lattice-gas models in statistical physics.

\begin{figure}[pt]
\centering
\includegraphics[width=0.3\textwidth]{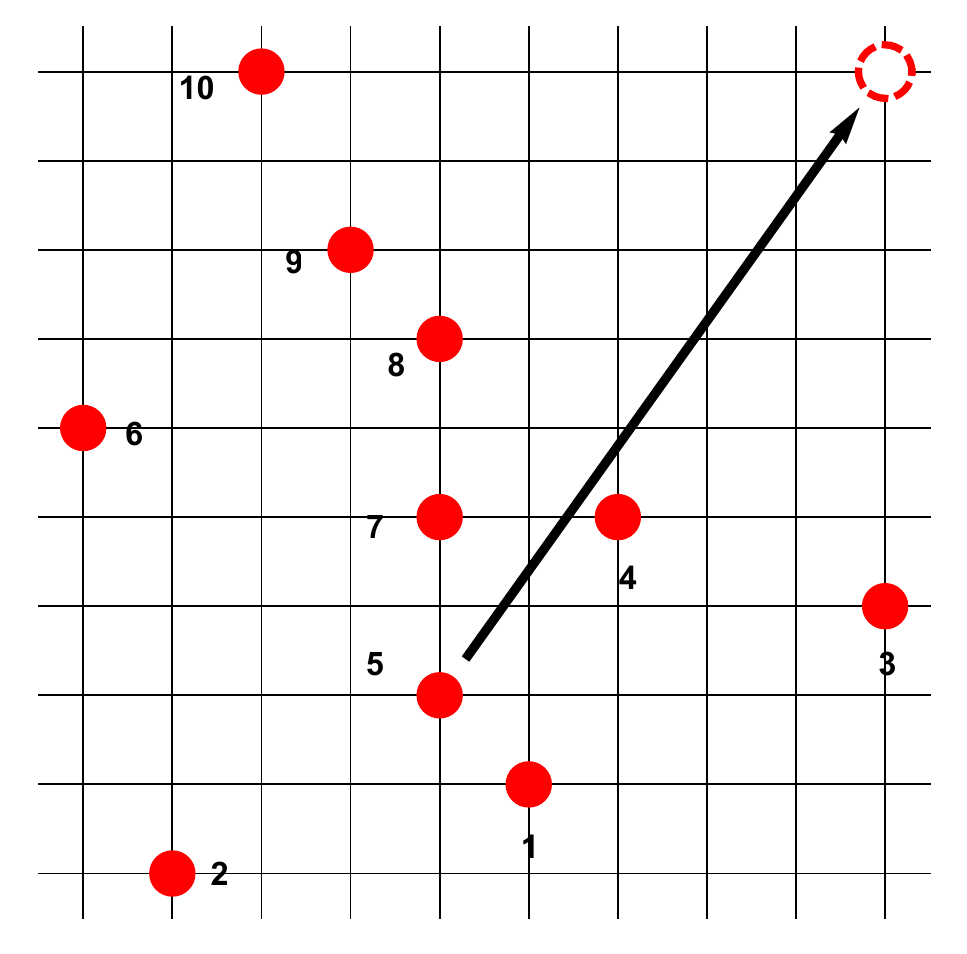}
\includegraphics[width=0.25\textwidth]{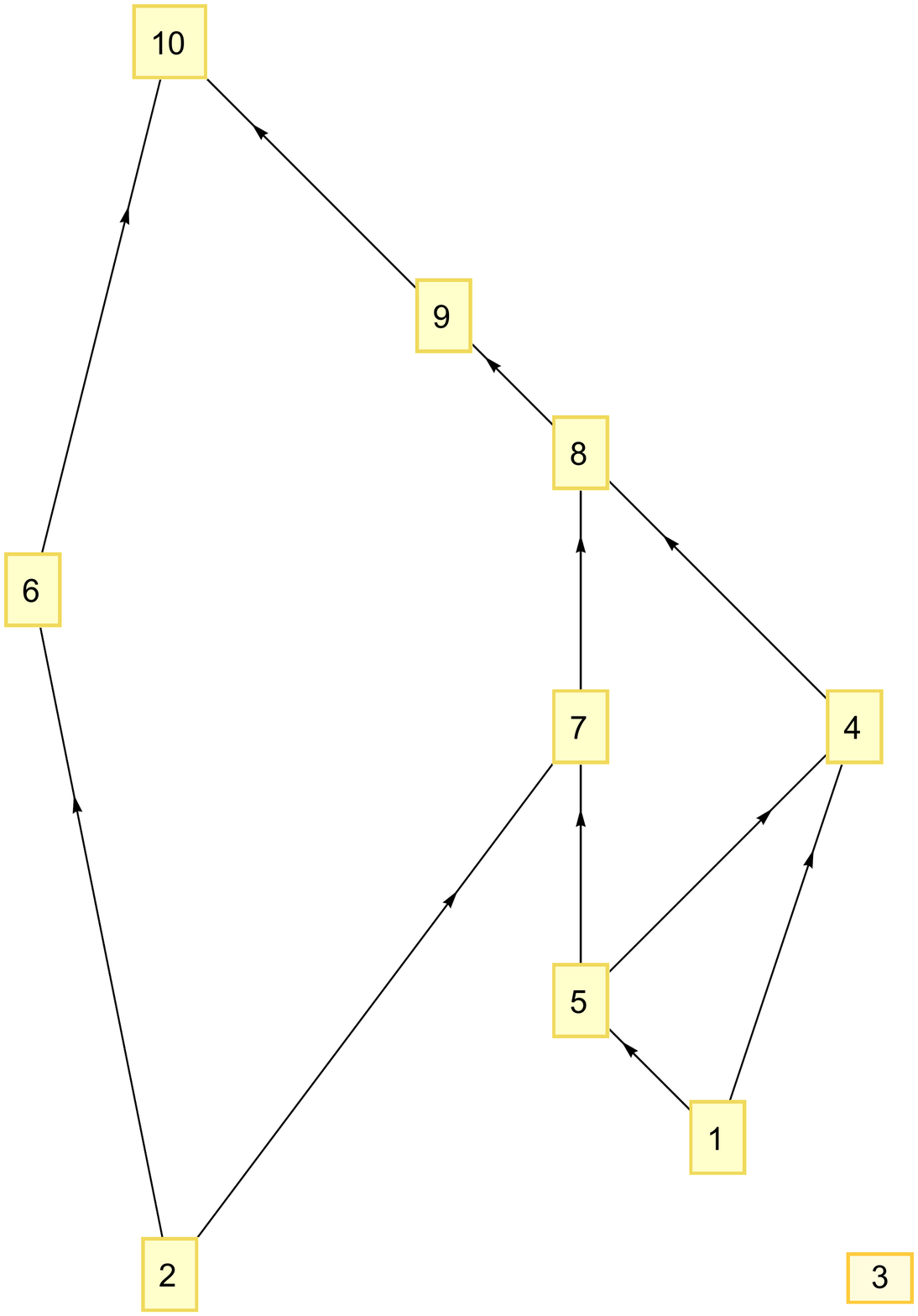}\hskip 0.4cm \includegraphics[width=0.32\textwidth]{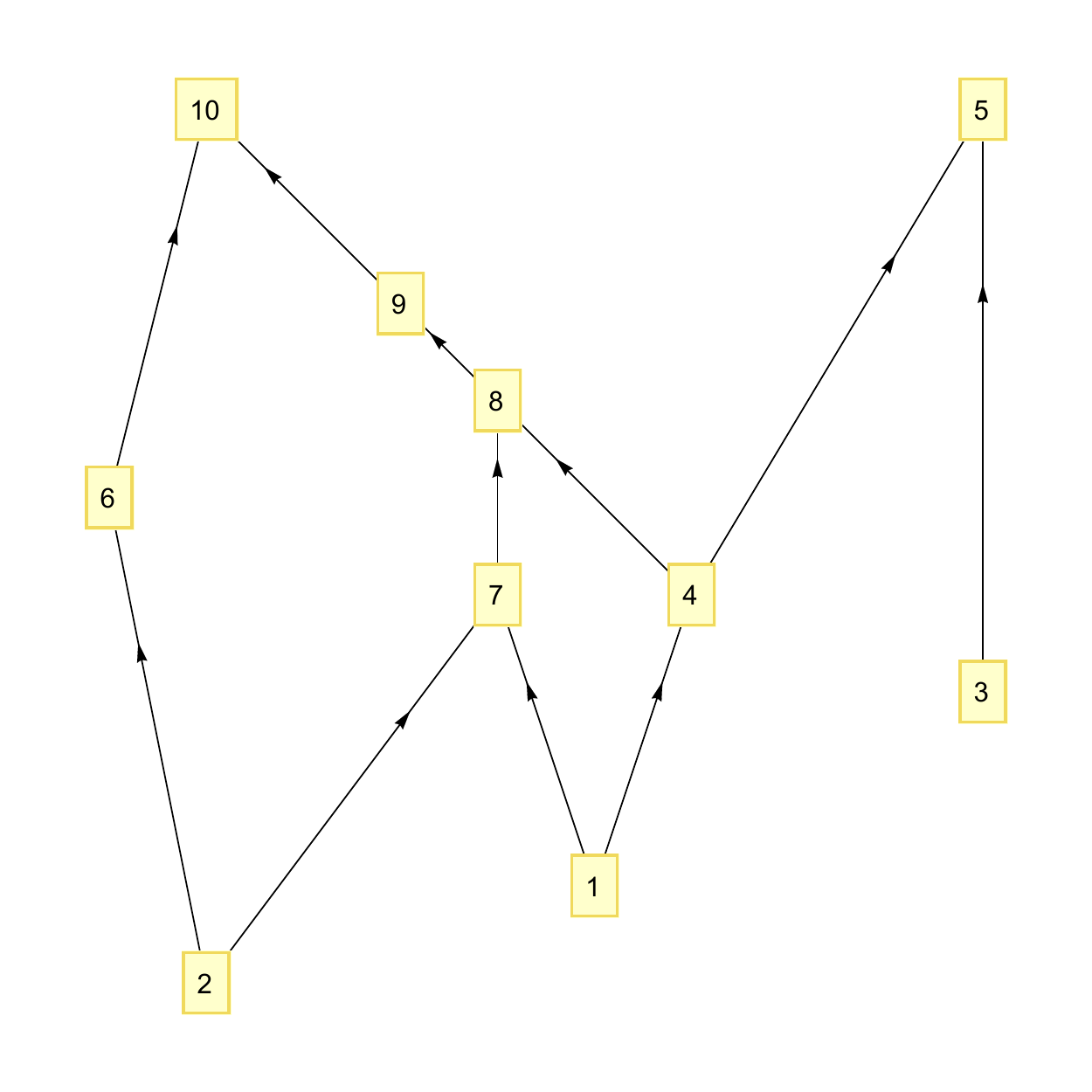}
\caption{An example of the \emph{lattice-gas move}. (i) A filled site is picked at random and moved to an unfilled site, 
  which itself is chosen at random  and independently. This move is typically non-local and can drastically
  change the associated causal set, shown here by  (ii) $\rightarrow$ (iii).}
\label{fig:move_example}
\includegraphics[width=\textwidth]{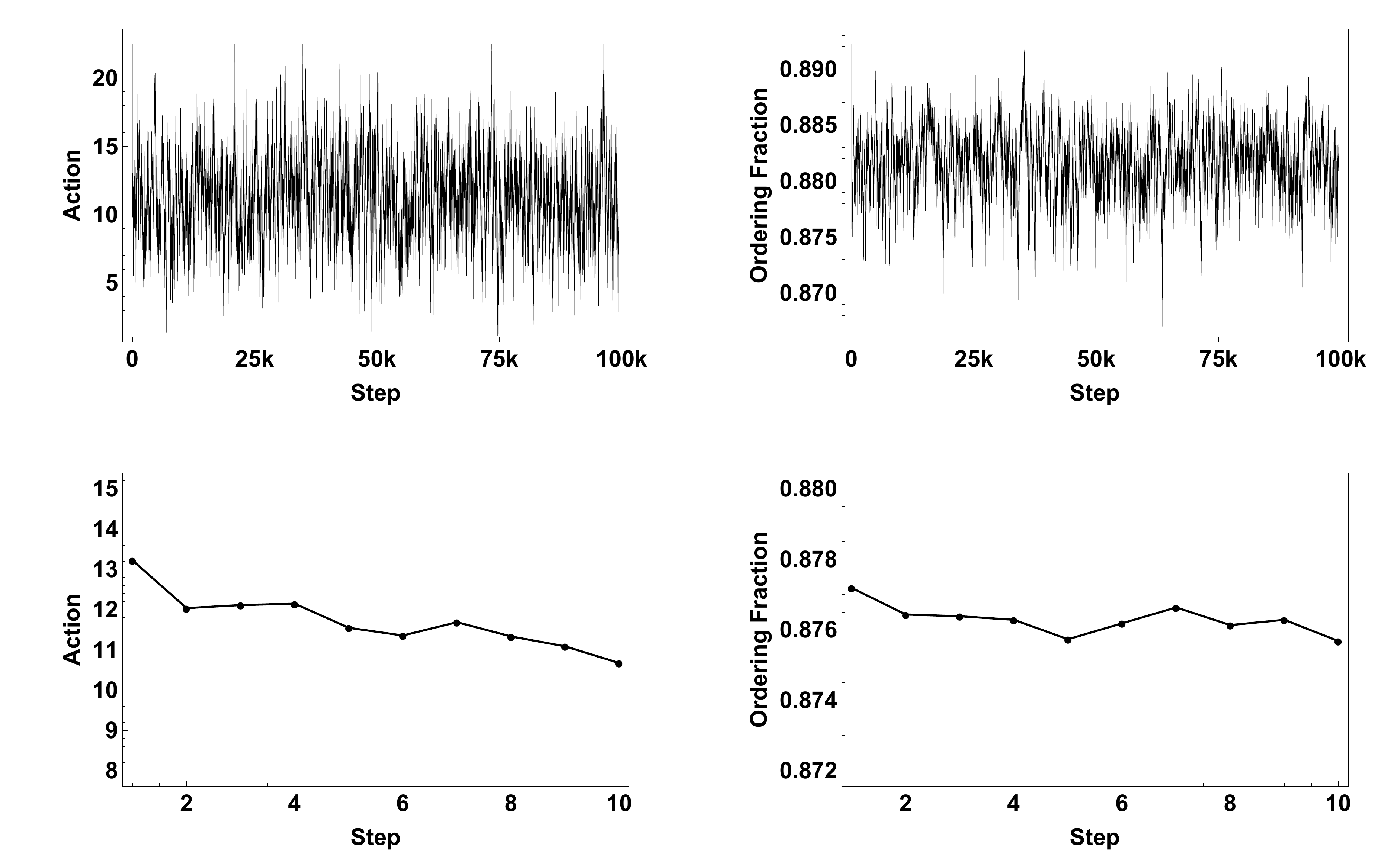}
\caption{The action and ordering fraction are plotted at each step in the Markov chain, with the former calculated using
  the non-locality scale $\epsilon=0.1$. These observables vary at each
  step which implies that a different causal set is generated at each step. The top two panels show the variation over
  $10^{5}$ steps while the bottom two panels show just the first $10$ steps.  These values were measured for $\dimm=2$ using $n=w=200$,
  $h=800$, after thermalising for $1000$ sweeps.}
\label{fig:perstep}
\end{figure}

Consider an $n$-site filling $E \in \cEcylmn_\dimm$ and let $c(E)$ be the associated causal set in $\ocylmn_\dimm$. Let $\bE$
represent the set of $m-n$ unfilled sites.  Pick one of the  filled
sites $e=(t,x)  \in E$ and one of the unfilled sites  $\bbe=(\bt, \bx) \in \bE$  independently and at random,   and swap
them, to get the new
filled site $e'=(\bt,\bx)$  and the new unfilled site   $\bbe'=(t,x)$.  One has moved from $E$ to a different
$n$-site filling $E' \in \cEcylmn_\dimm$ via the filled-empty exchange $e \leftrightarrow \bbe$. Fig.~\ref{fig:move_example} shows an example
of this lattice-gas move in $\dimm=2$. We can then use the standard 
MCMC  methods to  sample $\ocylmn_\dimm$ and calculate the expectation value of several order
invariants.  

In the above lattice-gas move, since $e$ and $\bbe$ are picked 
randomly in $E$ and $\bE$, the possibility of emptying the filled site $e$ to an empty site in its vicinity is
relatively small. Hence we expect that the resulting causal set $c'(E')$ will in fact be distinct from $c(E)$.  In
Fig.~ \ref{fig:perstep} we have shown the results of an MCMC  simulation\footnote{We will describe the details of these
  simulations in Sec.~\ref{sec:dynamics}.} for $\dimm=2$, $n=200$, $h=800,
w=200$, where the order invariants  (or observables) are the Benincasa-Dowker action  $\bds^{(2)}$ as well as the \emph{ordering
fraction}  plotted at
every MCMC step.
To ensure a large enough
lattice height $h$ for capturing the effects of non-trivial global spatial topology, a choice of aspect ratio
$\alpha=4$ is sufficient for our purposes. 
In Appendix~\ref{app:of} we calculate the expectation value of the
ordering fraction for a causal set that is approximated by the cylinder spacetime. For $\alpha=4$, this gives the value
of $\sim 0.88$, which is what we observe in Figs.~\ref{fig:perstep}, ~\ref{fig:2d_variable_volume_zerob}. 
 Even though two distinct causal sets can have the same values for these observables,  the converse is not
 true, i.e., if the two sets of observables are different, then so are the causal sets. Hence  it is clear that at almost every step one obtains a distinct causal set. 

\begin{figure}[!t]
\centering
\includegraphics[width=\textwidth]{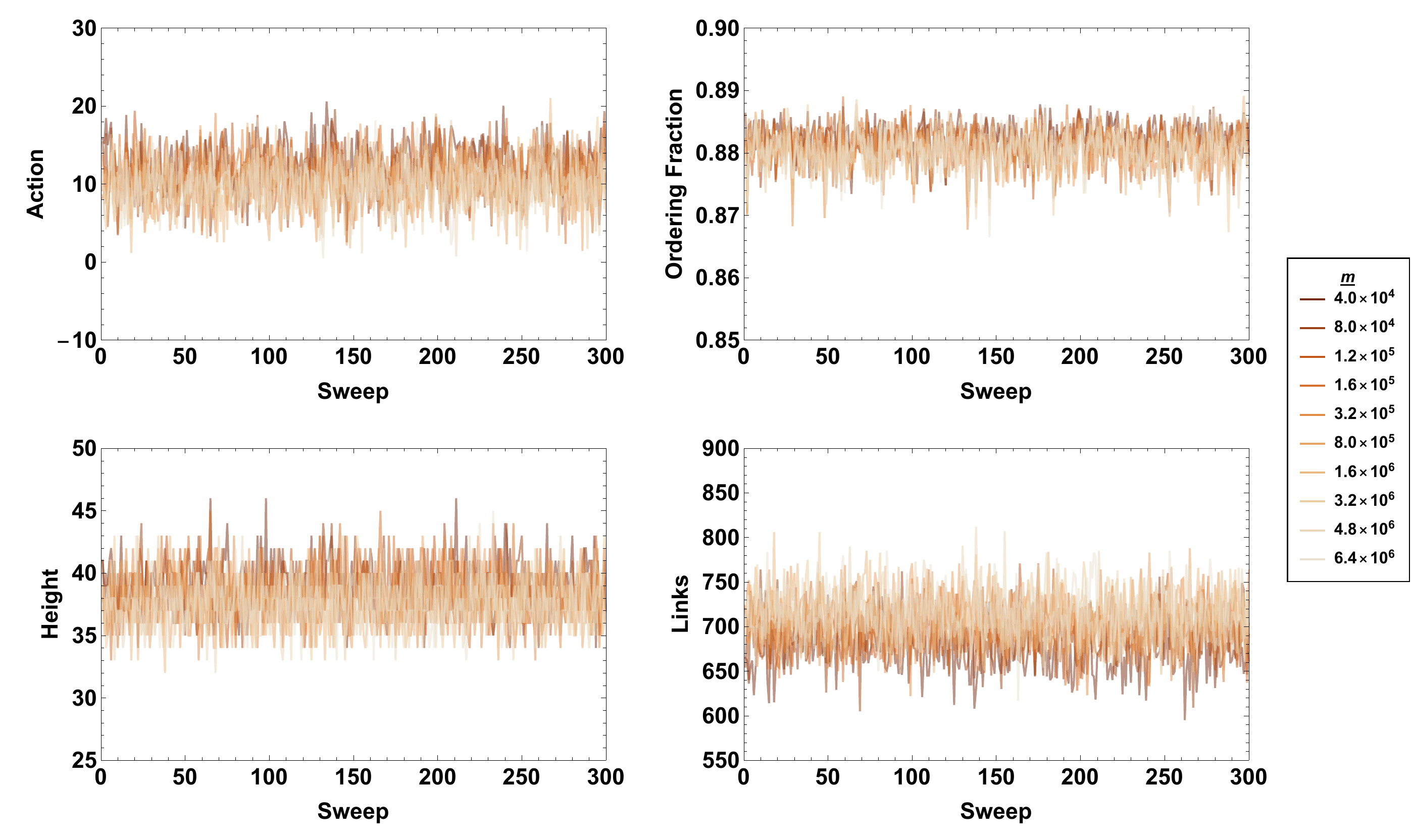}
\caption{Raw data for  the action, the ordering fraction, the height, and the number of
  links from lattice-gas  MCMC simulations which generate the typical $(m,n)$-orders, for   $\dimm=2$, $n=200$ and $\alpha=4$.
  When $m\gg n$, these order
  invariants can be seen to be independent of $m$.}
\label{fig:2d_variable_volume_zerob}
\end{figure}

To establish the asymptotic regime when $\Omega^{(m,n)}$ becomes independent of $m$, 
we run simulations for different values of $m$ while keeping $n$ and $\alpha$ fixed.  In
Fig.~\ref{fig:2d_variable_volume_zerob} we demonstrate this for $\dimm=2$, $n=200$, $\alpha=4$ for different filling
fractions $f=n/m$ or different values of $m=h\times w = \alpha \times w^2$. Here, instead of recording every step, we record every \emph{sweep}, defined as
$\binom{n}{2}$ steps. From the plotted data,  one sees a clear convergence of the thermalised configurations with
$m$. Hence for $\dimm=2$,  and these choices of $n$ and $\alpha$, $m=4\times 10^4$ is sufficiently in the asymptotic regime for our purposes.   

In the sample space $\Omega_n(M_\dimm,\eta)$, a special role is played by those causal sets that embed faithfully into, or
are \emph{approximated  by}  
$(M_\dimm,\eta)$.  An important question is how this class of causal sets is represented in $\ocylmn_\dimm$.   

\begin{figure}[!t]
\centering
\includegraphics[width=0.5\textwidth]{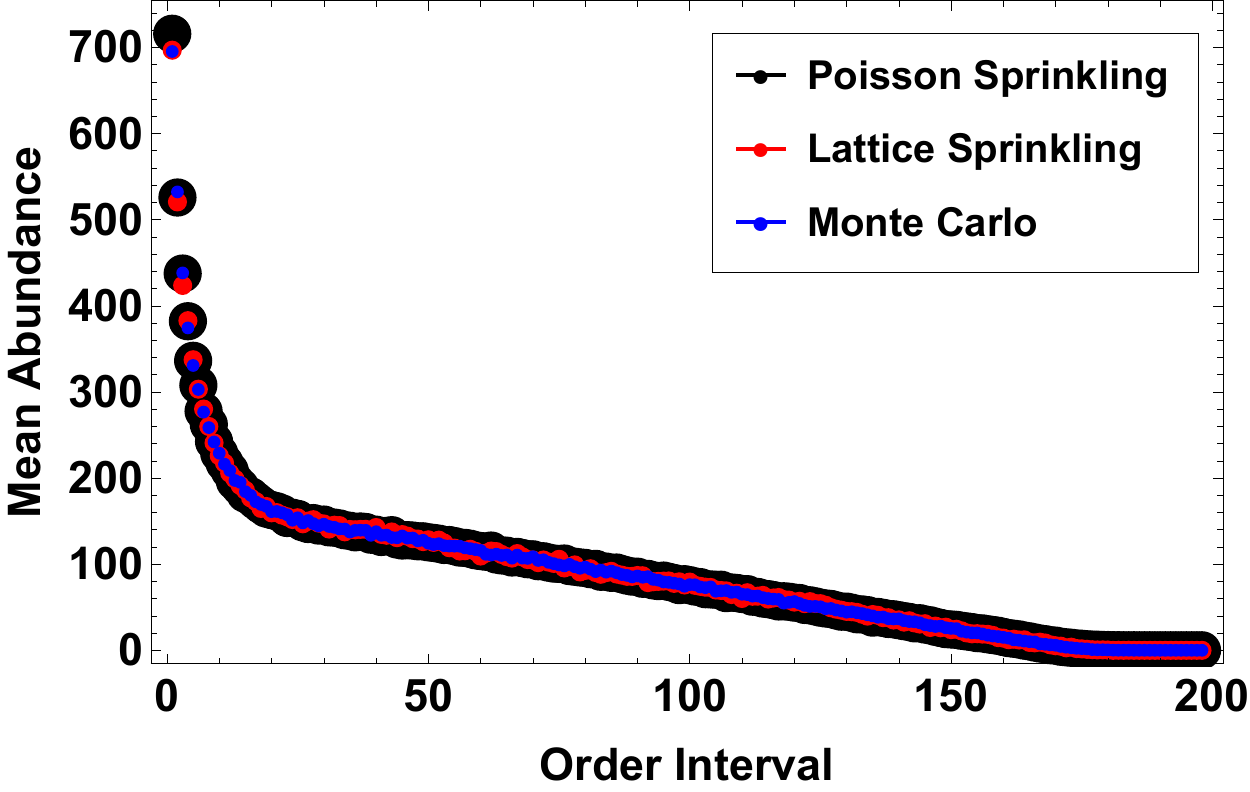}%
\includegraphics[width=0.5\textwidth]{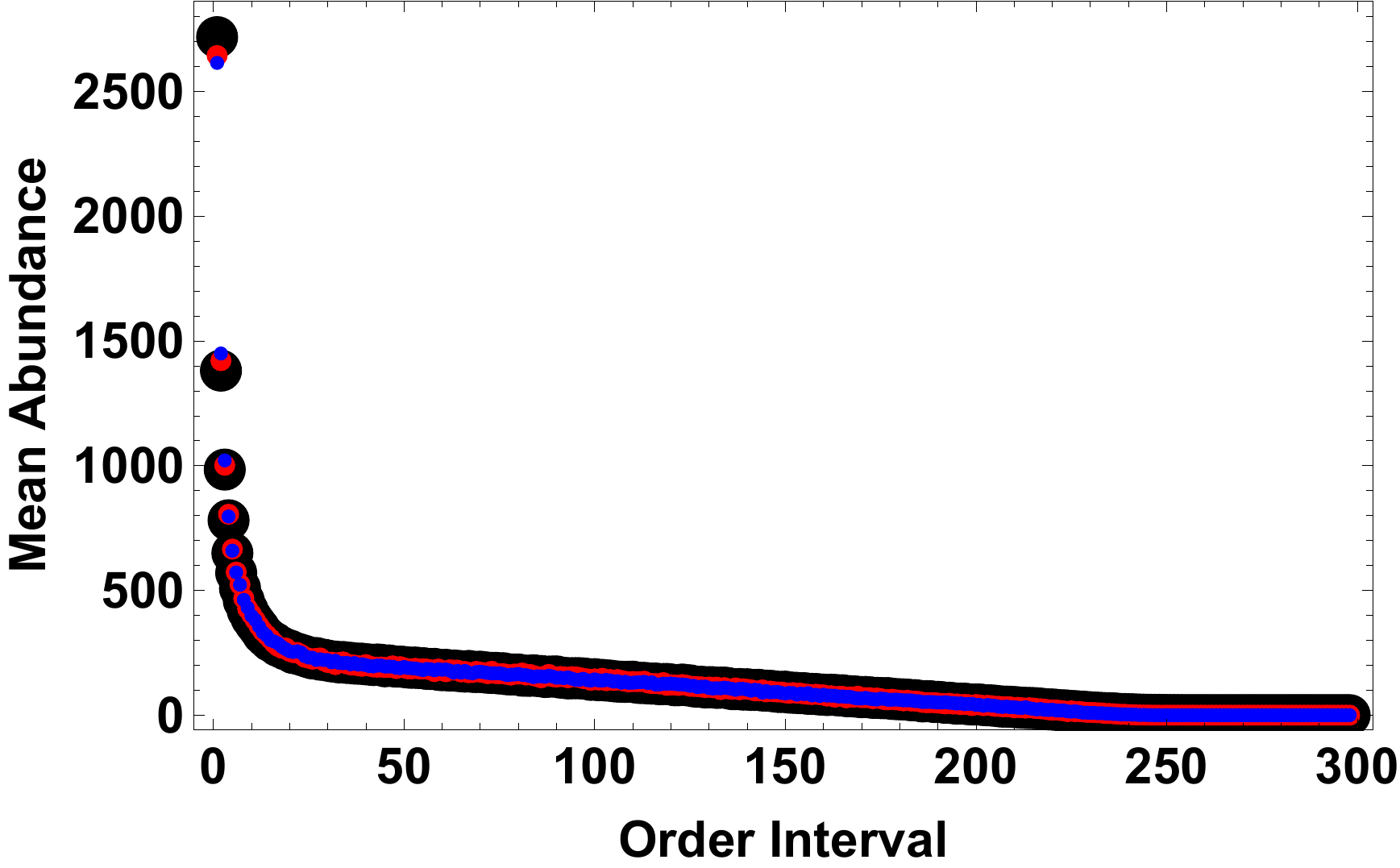}
\caption{Mean interval abundance curves for $\dimm=2$ (left) and $\dimm=3$ (right). The three curves represent a Poisson sprinkling
  into the continuum, a random $(m,n)$-order sprinkled onto the lattice, and a typical $(m,n)$-order generated via
  a lattice-gas MCMC simulation.}
\label{fig:sprinkling_comparisons}
\end{figure}

It was shown in \cite{winkler,es} that the typical $2$-order is a  \emph{random $2$-order} obtained by  choosing the
two linear orders (or light-cone $(u,v)$-coordinates)  randomly and independently. It was then shown in \cite{2dorders} that the class of  random orders are
those that faithfully embed into $(D_2,\eta)$.   In analogy, we define   the \emph{random $(m,n)$-order}  to be that
obtained from a random filling of $\mlcylm_\dimm$.  Are these typical? 
 We can answer this question at least partially by employing the MCMC algorithm on $\cEcylmn_\dimm$, and comparing the expectation value of various 
 observables with that of a random $(m,n)$-order.  As we see from Figs.~\ref{fig:sprinkling_comparisons} where we have plotted the expectation
 values of the \emph{abundance
 of $r$-element intervals},  one cannot distinguish a  typical causal set in
$\ocylmn_\dimm$ from a  random $(m,n)$-order either in $\dimm=2$ or $\dimm=3$.  We find the same when further comparing with
the causal sets obtained from Poisson sprinklings into $(M_\dimm,\eta)$, Eqn.~\ref{eq:cylspt}.  

\section{Setting up the Dynamics}
\label{sec:dynamics} 

Given  $\Omega_n(M_\dimm,\eta)$,  we can write the dimensionally restricted  path-sum for causal set quantum gravity as 
\begin{equation}
Z_n ^{(\dimm)}=\sum_{c \in \Omega_\dimm(M_\dimm,\eta)} \exp(i\bds^{(\dimm)}(c)/\hbar). 
\end{equation} 
In order to evaluate the expectation values of 
covariant observables or order-invariants on the computer, we first introduce an analytic continuation
parameter $\beta$  such that  \cite{2dqg} 
\begin{equation}
Z_n ^{(\dimm)} (\beta)=\sum_{c \in \Omega_n(M,g)} \exp(i\beta \bds^{(\dimm)}(c)/\hbar) \, \, \underset{\beta \to  i \beta}{\longrightarrow}   \, \,
\cZ_n ^{(\dimm)} (\beta)=\sum_{c \in \Omega_n(M,g)} \exp(-\beta\bds^{(\dimm)}(c)/\hbar). 
  \end{equation} 
 This converts the quantum path-sum $Z_n ^{(\dimm)} (\beta)$ to a statistical partition function $\cZ_n ^{(\dimm)} (\beta)$ over discrete Lorentzian geometries, which we
 can try to simulate on the computer. 

 The latticisation of $(M_\dimm,\eta)$ into $\mlcylm_\dimm$ with the associated sample space
 $\ocylmn_\dimm$  is the  first step in this implementation,  using which we define the discrete partition
 function
 \begin{equation}
   \cZ_{n,m}^{(\dimm)}(\beta) \equiv \sum_{c \in \ocylmn_\dimm} \exp(-\beta\bds^{(\dimm)}(c)/\hbar).
   \label{eq:partn_stat} 
 \end{equation}
 We can then implement the MCMC lattice-gas  move described above to perform a weighted walk through $\ocylmn_\dimm$, using the Metropolis algorithm, with the expectation values $\av{O}$  of
different {order-invariants} $O$ being  calculated using importance sampling.  

In simulations, the lattice $\mathcal{L}^{(m)}$ is represented by the integer sequence $L=\{0,\ldots,m-1\}$, where the
entry $L_i$ represents an element $e$ with spacetime coordinates $t_i=\left\lfloor L_i/w\right\rfloor/h$ and
$\theta_i=(2\pi/w)(L_i\mod w)$\footnote{Here the index $i$ refers to the lattice point and not the direction.}. To
represent an $(m,n)$-order $E\in\mathcal{E}_\dimm^{(m,n)}$ on $\mathcal{L}^{(m)}_\dimm$, we select $n$ unique entries
from $L$, or equivalently, the first $n$ entries of a random permutation of $L$.  Thus, 
a random permutation of $L$ yields a random $(m,n)$-order. Moreover, a swap between a  randomly chosen element in the first $n$
entries with a randomly and independently chosen element from the last
$m-n$ entries, represents  the lattice-gas move $e\leftrightarrow\bar{e}$. Because the lattice-gas move is typically non-local,
every time we update $L$ we must also recalculate all causal relations using the spacetime coordinates
$(t_i,\theta_i)$. The representation of the causal (adjacency) matrix for the corresponding causal set, as well as the
implementation of the action calculation, are described in~\cite{cunningham2018causal}.

States are updated in the Markov chain according to the Metropolis rule: a new configuration resulting from a lattice-gas move is accepted if
$\exp(-\beta\Delta S_{BD}^{(\dimm)})>\mathfrak{u}$, where $\mathfrak{u}$ is a uniform random variable in the range
$[0,1)$. Observables are recorded each sweep, or $\binom{n}{2}$ steps, and simulations are run for at least $10^4$
sweeps. Despite the compact data structures limiting memory usage ($\sim 10$ MB) and parallel algorithms used to do
calculations, it is prohibitively expensive to increase $n$ above $O(10^2)$ because each of the $10^4\binom{n}{2}$ updates still requires $\sim 100\,\mu\mathrm{sec}$ and the simulation has an overall complexity of $O(n^5)$.

For this reason, we have chosen to study $n=200,300$ in order to limit individual experiments to finish in under three days when running on machines with up to 56 cores. Due to the exceptionally high volume of data needed to understand the large space of $(m,n)$-orders in two and three dimensions, we ran these simulations on supercomputers at Northeastern University (Discovery cluster), Raman Research Institute, Universit\'e de Sherbrooke (MP2B cluster), and Simon Fraser University (Cedar cluster) over approximately 283.6 core-years. At a future date, the code used to perform these experiments will be merged with the Causal Set Generator~\cite{cunningham2017generator} and the data will be published in the Encyclopedia of Quantum Geometries~\cite{eqg2019}.

\section{\texorpdfstring{Causal set dynamics in $\dimm=2,3$.}{Causal set dynamics in d=2,3}}
\label{sec:2d3d} 

We now present the results from  simulations in  $\dimm=2$ and $\dimm=3$ for the choice spacetimes, 
Eqn.~\ref{eq:cylspt}. While we have performed
simulations over several values of $n$ and non-locality parameter $\epsilon$ (see Eqn.~\ref{eq:bdactions} and \ref{eq:fnes}), for the purpose of this work we fix $\epsilon=0.1$ and focus on the largest
values of $n $ for which the simulations thermalise in reasonable time: $n=200$ for $\dimm=2$ and $n=300$ for
$\dimm=3$.  We choose $\alpha=4$, and $w=n^{1/(\dimm-1)}$, so that $m=4n^2 \sim 1.6\times 10^5$ in $\dimm=2$ and $m=4n^{3/2} \sim 2.0\times 10^4$ in $\dimm=3$.

\begin{figure}[!t]
\centering
\includegraphics[width=\textwidth]{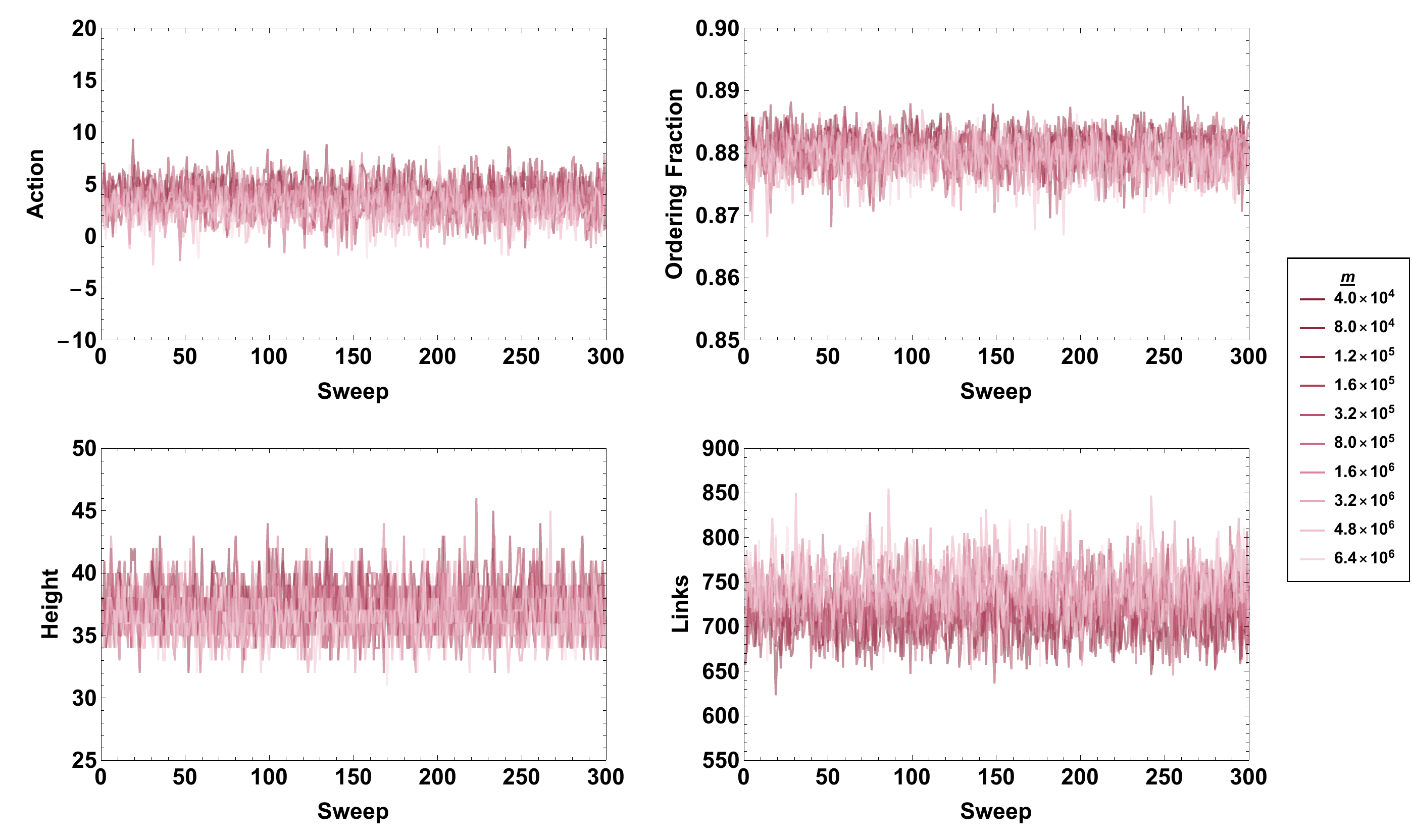}
\includegraphics[width=\textwidth]{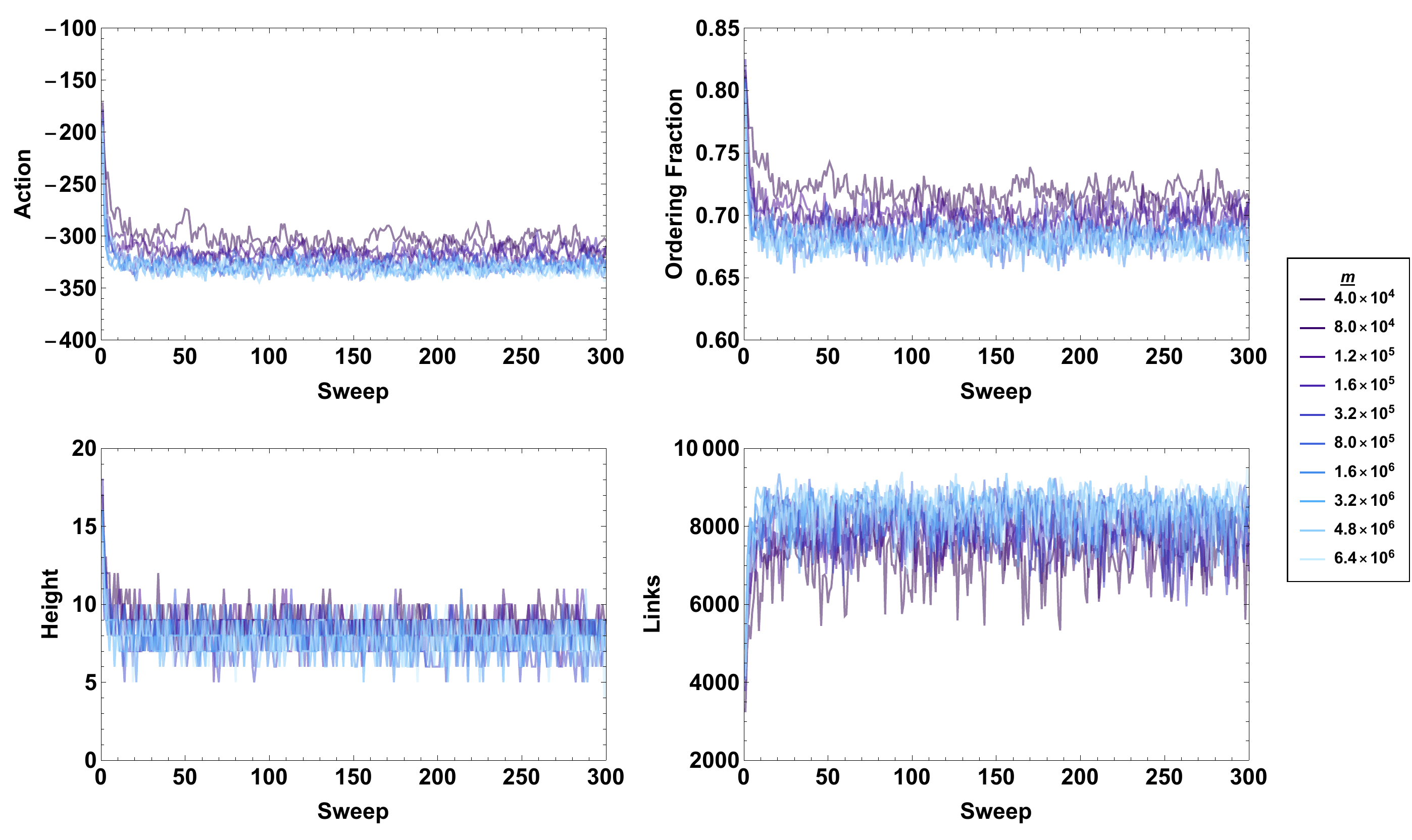}
\caption{Raw data for the action, the ordering fraction, the height, and the number of
  links for $\dimm=2$, $n=200$, $\alpha=4$, $\epsilon=0.1$ and $\beta=0.8$ (upper four panels) and $\beta=3.2$ (lower four
  panels).}
\label{fig:2d_variable_volume}
\end{figure}

\begin{figure}[pt]
\centering
\includegraphics[width=\textwidth]{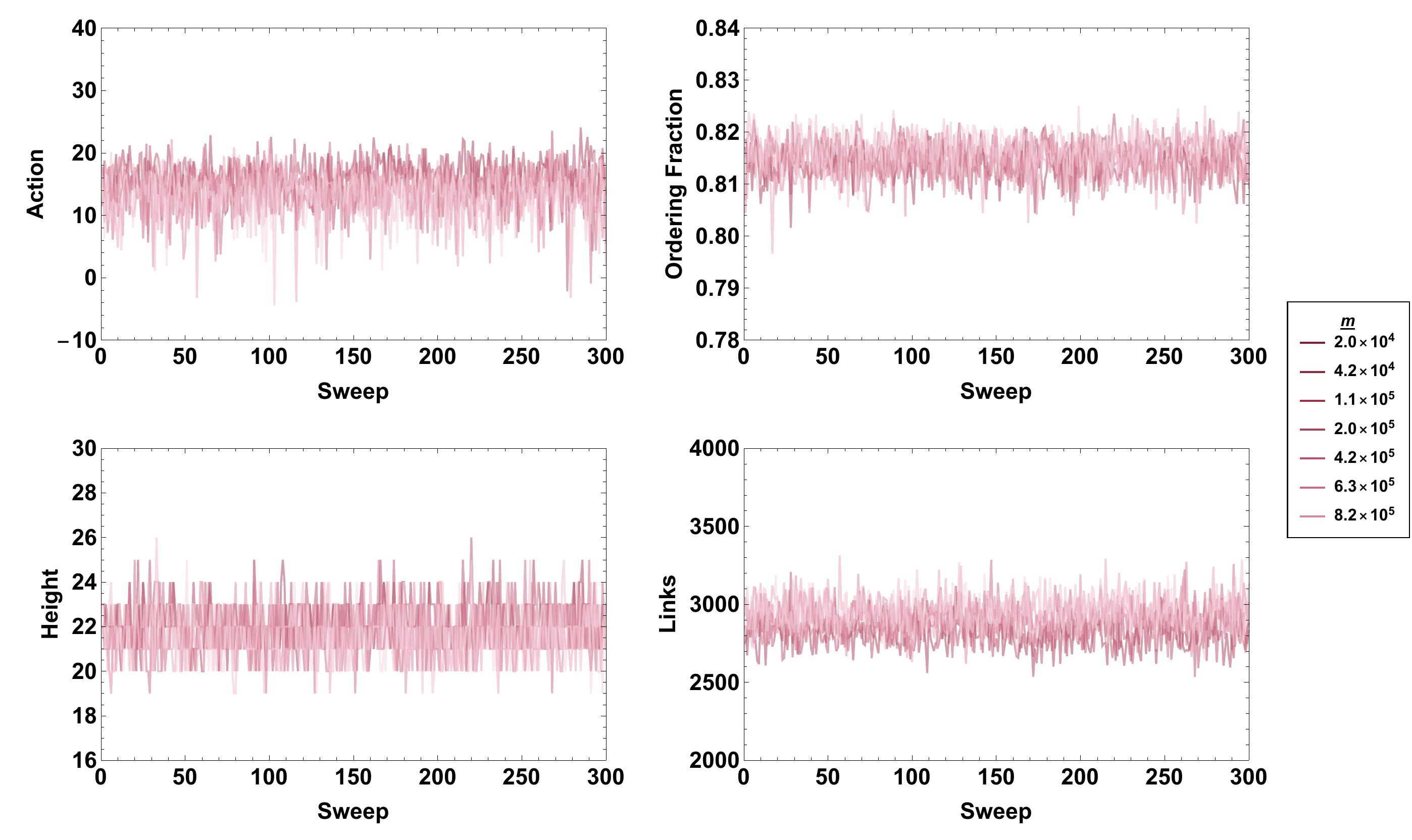}
\includegraphics[width=\textwidth]{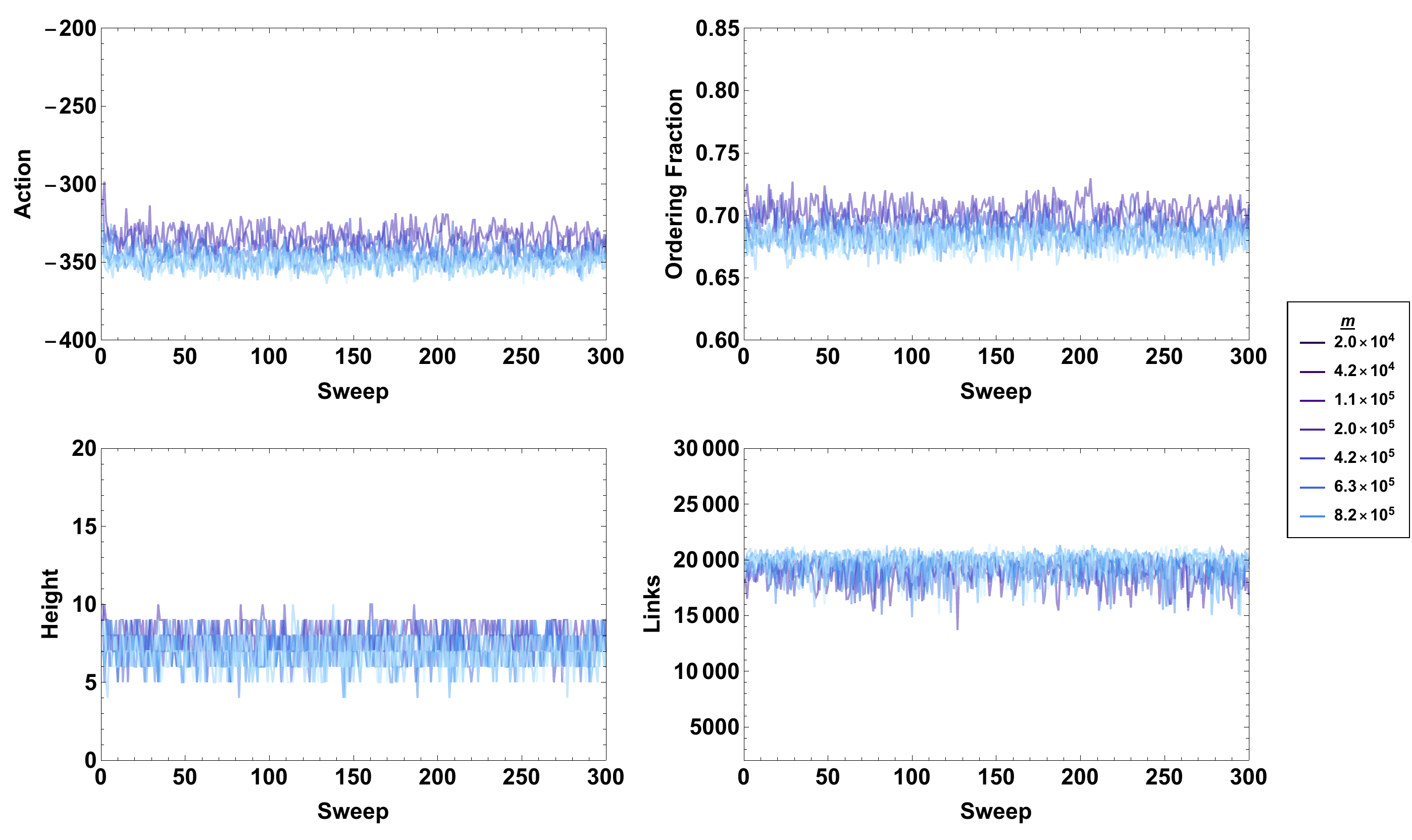}
\caption{Raw data for the action, the ordering fraction, the height, and the number of
  links for $\dimm=3$, $n=300$, $\alpha=4$, $\epsilon=0.1$, and $\beta=1.0$ (upper four panels) and $\beta=3.0$ (lower four panels).}
\label{fig:3d_variable_volume}
\end{figure}

To verify if our choice of $m$ is large enough,  in each dimension we vary over different values of $m$ and run the MCMC
simulation for two choices of $\beta\neq0$. We  find that the expectation values of the observables are roughly 
independent of $m$, as shown in  Figs.~\ref{fig:2d_variable_volume},~\ref{fig:3d_variable_volume}, suggesting
that the system is in the desired asymptotic regime.  

While the simulations can be carried out for larger values of $n$ while fixing  $\beta$,  the  focus here is to
study a range of $\beta$ to scan  for phase
transitions.
We have generated data for several observables including the BD
action, the ordering fraction, the \emph{height} and the number of \emph{links}.  In the following analysis, we consider the raw data or the expectation value of these four observables. We have also computed the
\emph{abundance}  $N_i$ of the
\emph{$(i-2)$-element intervals}  in all cases, but have not exhibited their change as a function of $\beta$
explicitly, since there are too many of them to show. We will however use them to characterise the typical causal sets
in the two phases in what follows.   At every sweep, we also  save the actual
configuration on the lattice, so that we can record the changes in the configurations. 

\begin{figure}[t]
\centering
\includegraphics[width=\textwidth]{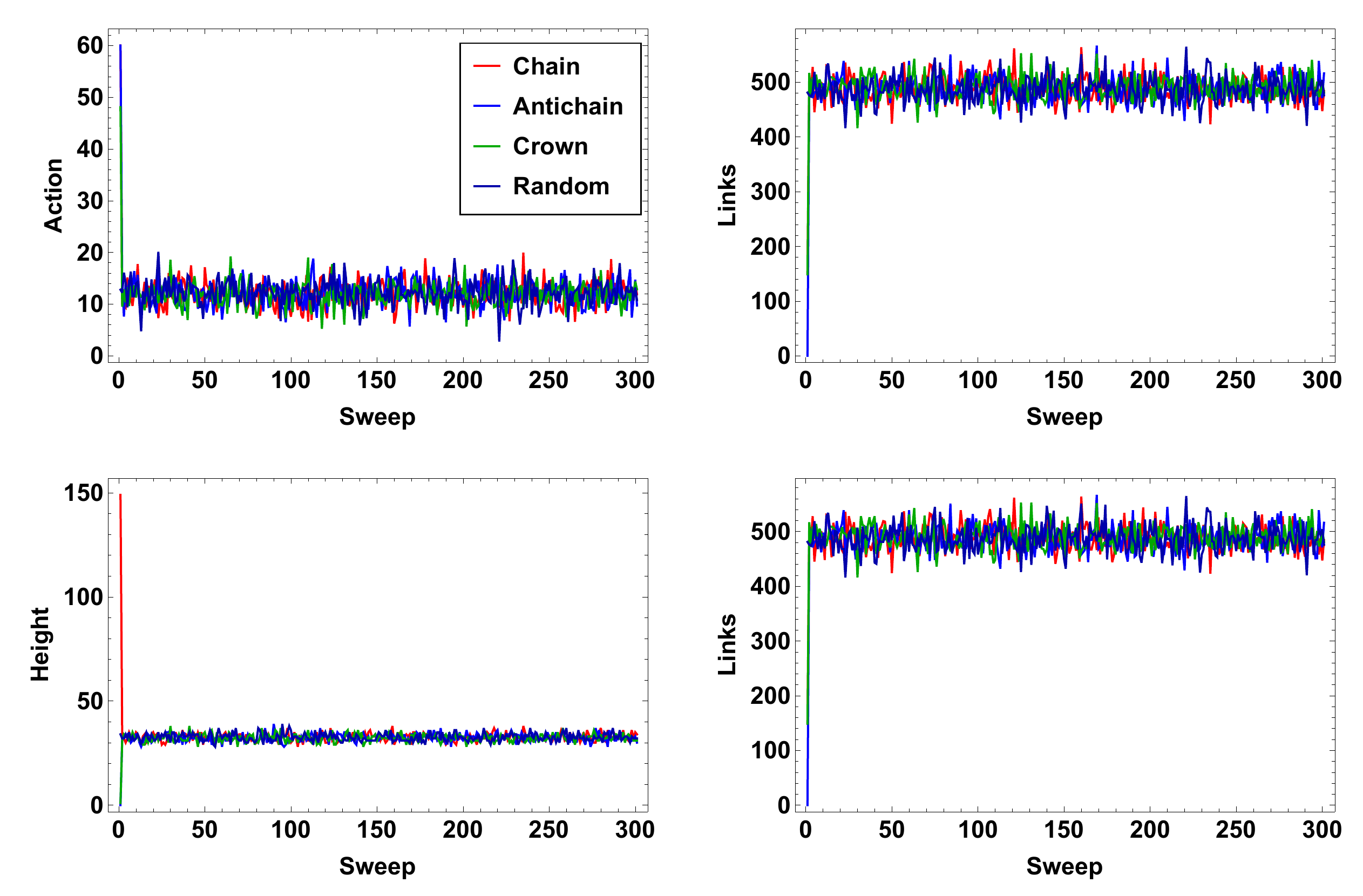}
\caption{The observables per sweep for $\dimm=2$, $n=150$, $\epsilon=0.1$, $\beta=0.2$, and $\alpha=4$ for different initial conditions.}
\label{fig:ergodic}
\end{figure}

\begin{figure}[pt]
\centering
\includegraphics[width=\textwidth]{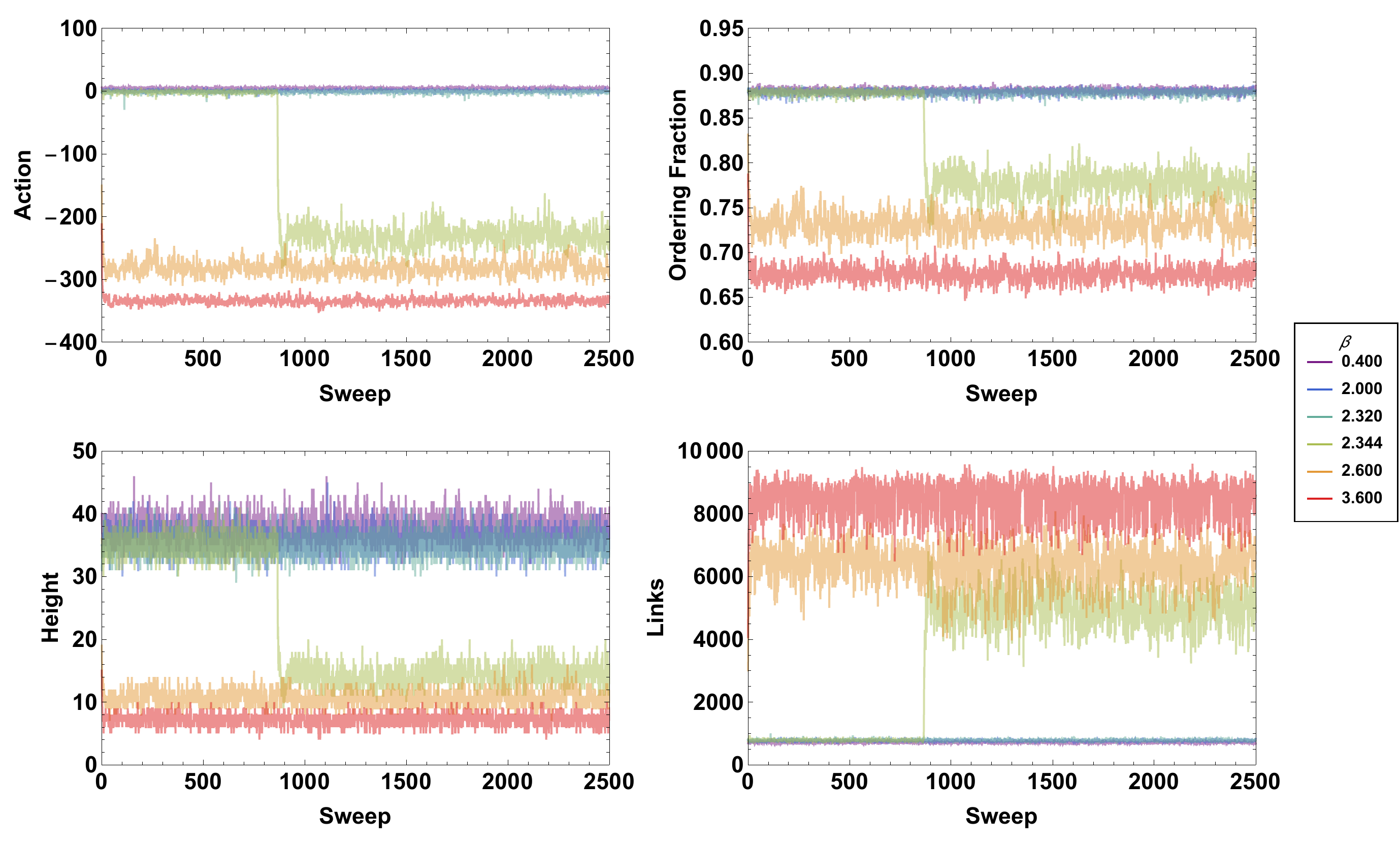}
\includegraphics[width=\textwidth]{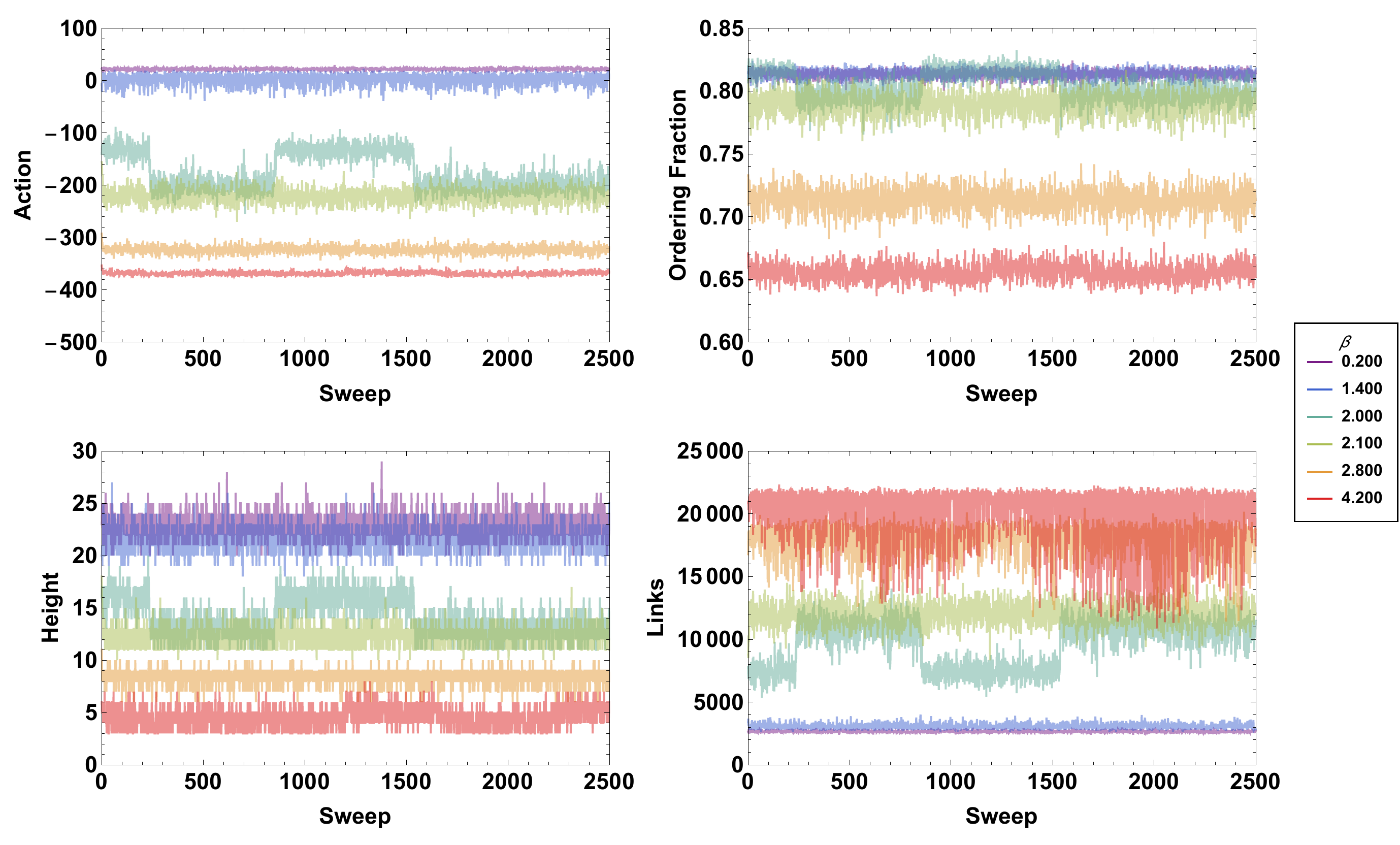}
\caption{Raw data for the action, the ordering fraction, the height, and the number of links for $\epsilon=0.1$ and $\dimm=2$, $n=200$ (upper four panels) and $\dimm=3$, $n=300$ (lower four panels).}
\label{fig:rawdata}
\end{figure}

To begin, we check for ergodicity by choosing different initial configurations and find that the thermalised
configurations are independent of this choice.  In Fig.~\ref{fig:ergodic}, we show a manifestation of this ergodicity in
$\dimm=2$. The choices include the completely ordered $n$-element set, or \emph{chain}, the  set of unrelated
$n$ elements, or \emph{antichain}, the random $(m,n)$-order described in Sec.\ref{sec:sample}, and the \emph{crown poset}. As is
evident, the thermalised value of the observables are independent of the initial configuration. 

Next, we present the raw data for  various observables for $\dimm=2,3$ for different values of
$\beta$ in Fig.~\ref{fig:rawdata}.  These figures demonstrate very clearly that the
configurations are thermalised  at different values of  $\beta$, {\it except} in a range of  intermediate values.  There
is a  clustering of  the observables for 
small $\beta$ (``hot'' configurations) and  for large beta (``cold''
  configurations). In the intermediate range, however 
the observables fluctuate strongly. In $\dimm=2$, Fig.~\ref{fig:rawdata},  we see that for  $\beta = 2.344 $, the observables oscillate between
two values, while in $\dimm=3$, this happens around $\beta = 2.0$. These fluctuations are
characteristic of a first order phase transition and resemble the behaviour of  the $2$-orders \cite{2dqg,fss}. In
$\dimm=2$, just by eyeballing Fig.~\ref{fig:rawdata}, we note that  the ordering fraction for the hot phase remains
close to the value of $0.88$ expected from a causal set that is  approximated by the flat cylinder spacetime.  Thus, even
without further analysis, the data is clearly consistent with a manifold-like phase for small $\beta$. As we will see
this is indeed the case. 

\begin{figure}[t]
\centering
\includegraphics[width=\textwidth]{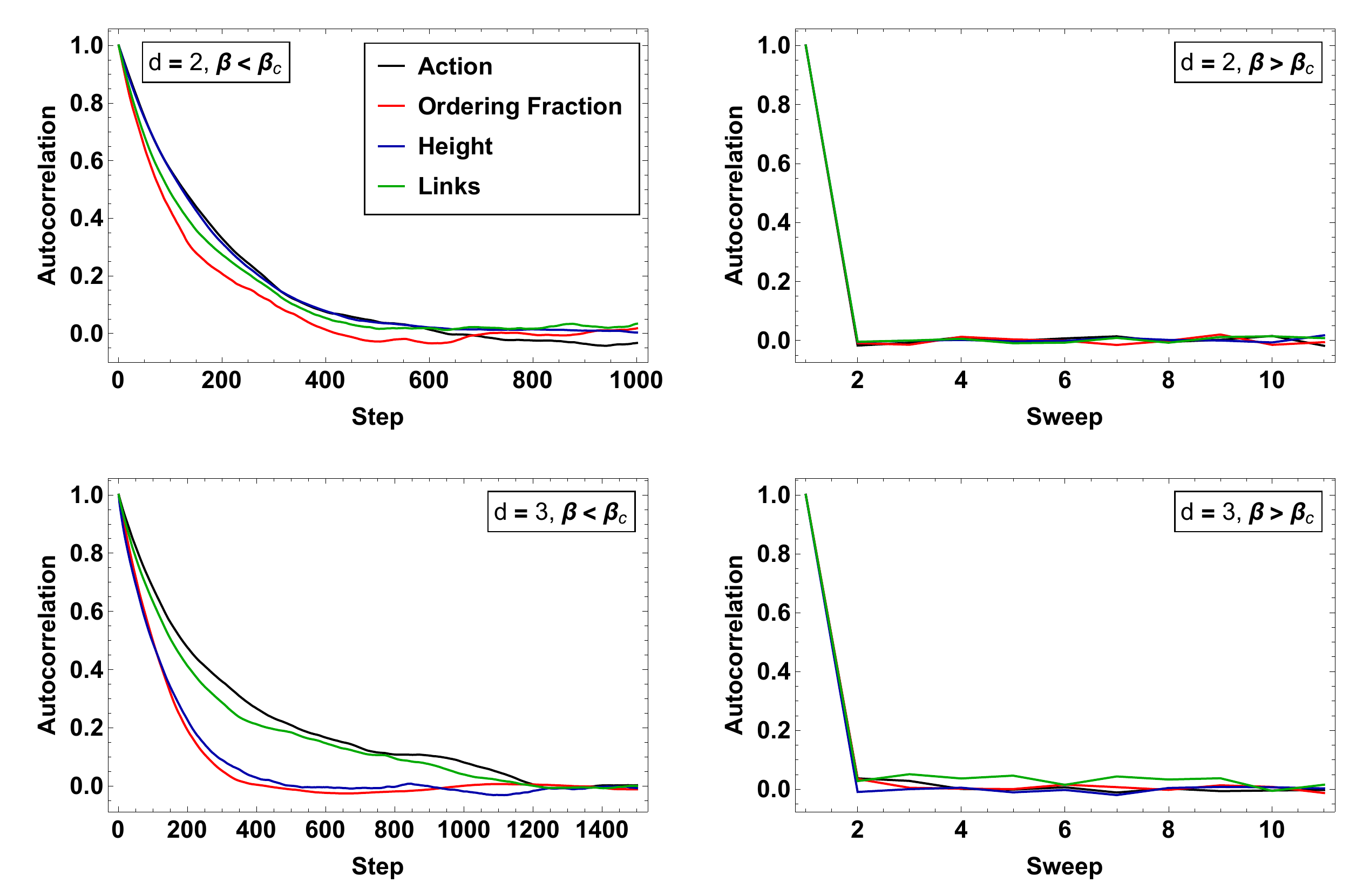}
\caption{Examples of autocorrelation function for various observables for the hot ($\beta<\beta_c^{(\dimm)}$)  and cold
  ($\beta >\bcd$)  phases in $\dimm=2,3$.}
\label{fig:autocorr}
\end{figure}

\begin{figure}[pt]
\centering
\includegraphics[width=0.95\textwidth]{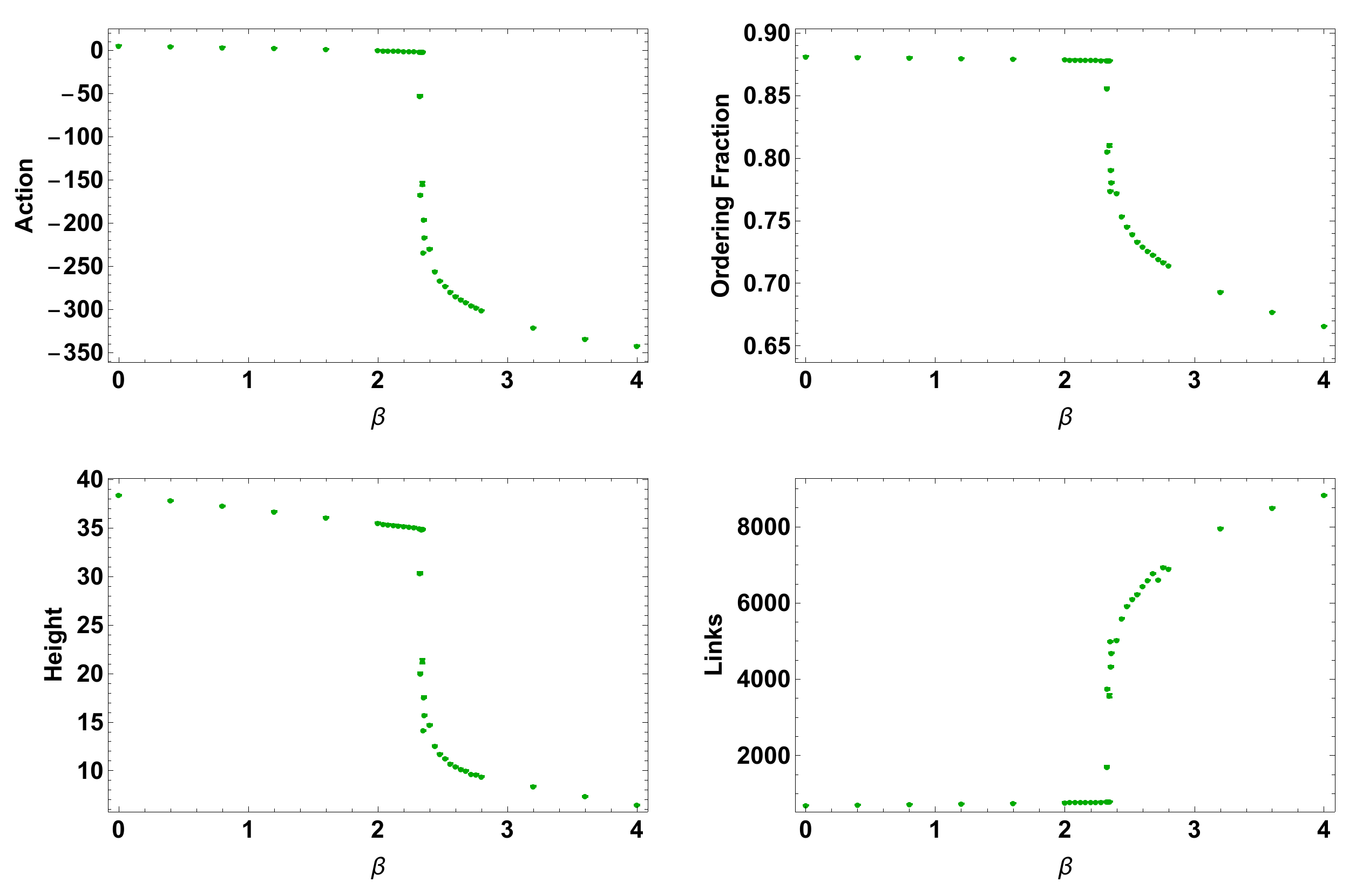}
\includegraphics[width=0.95\textwidth]{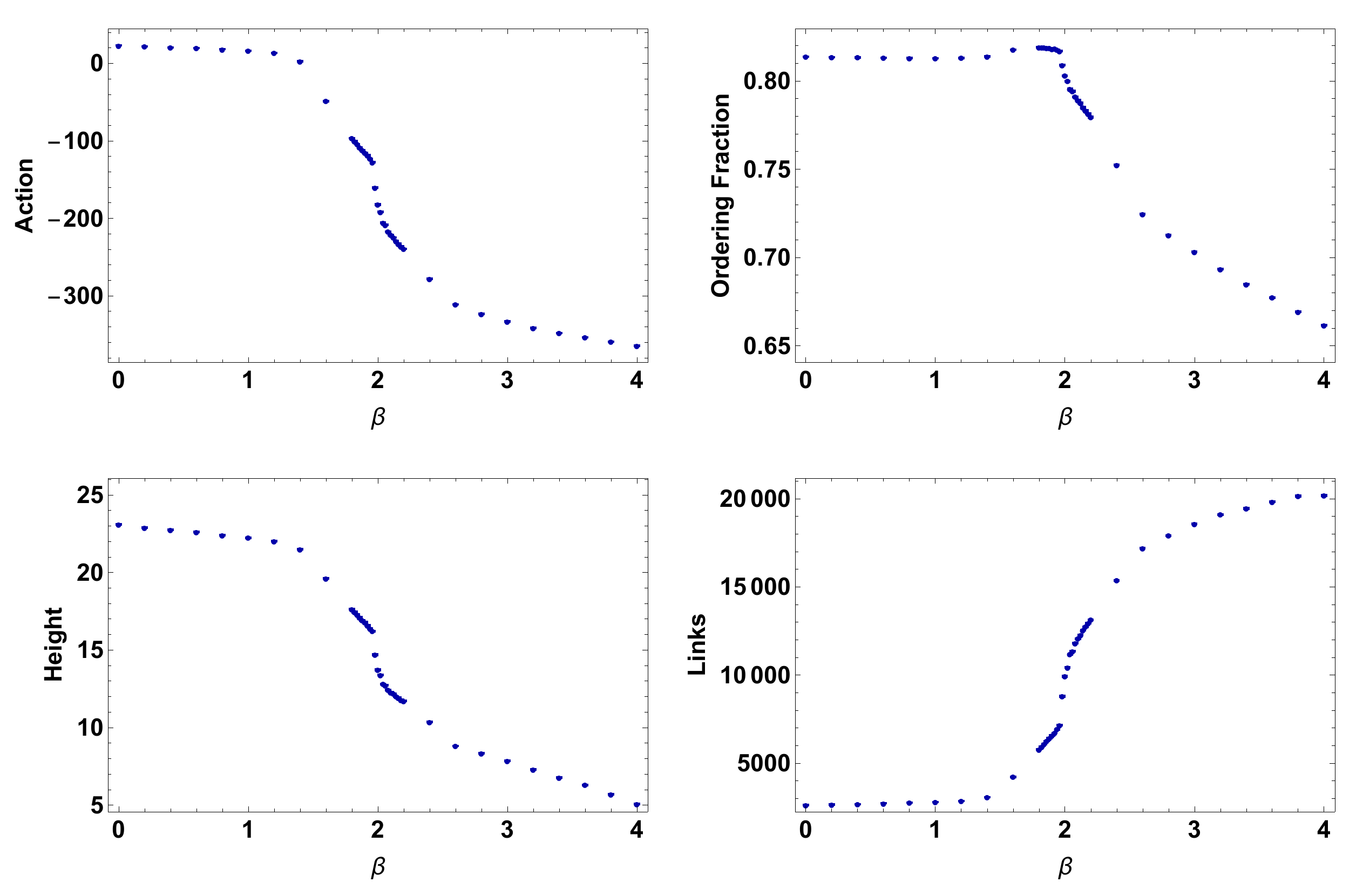}
\caption{The expectation value of various observables as a function of $\beta$ for $\dimm=2$ (upper four panels) and $\dimm=3$ (lower four panels). The error bars are shown but are very small.}
\label{fig:observables}
\end{figure}

\begin{figure}[pt]
  \centering
  \includegraphics[width=0.7\textwidth]{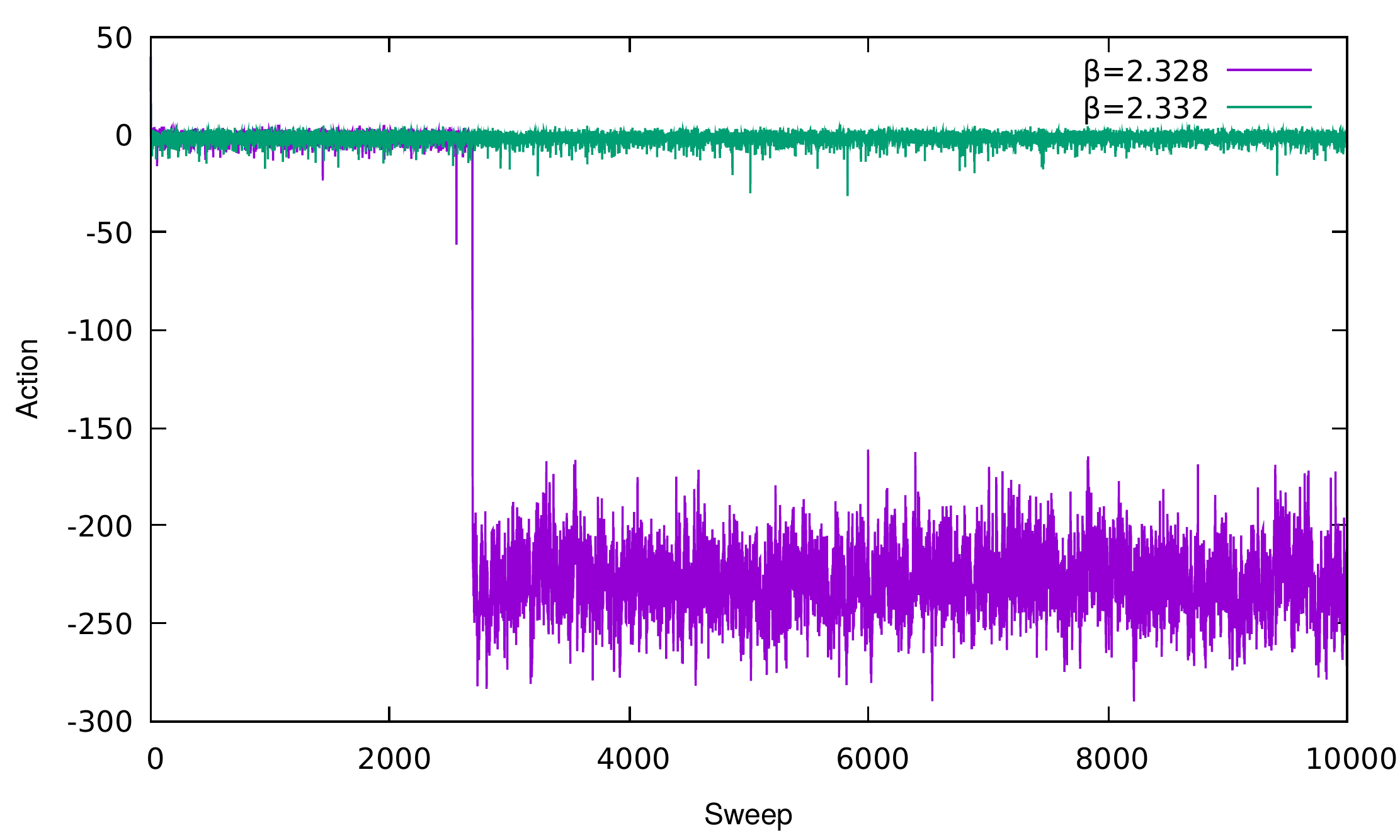}
  \caption{Raw data for the action in $\dimm=2$, showing the poor thermalisation near $\bctwo$ for $10^4$ sweeps. The system spends a
    longer time in the hotter phase at a lower temperature $\beta^{-1}=2.332^{-1}$  than at the higher temperature
    $\beta^{-1}=2.328^{-1}$. }\label{fig:badthermalisation} 
\end{figure}

\begin{figure}[pt]
  \centering
  \includegraphics[width=0.8\textwidth]{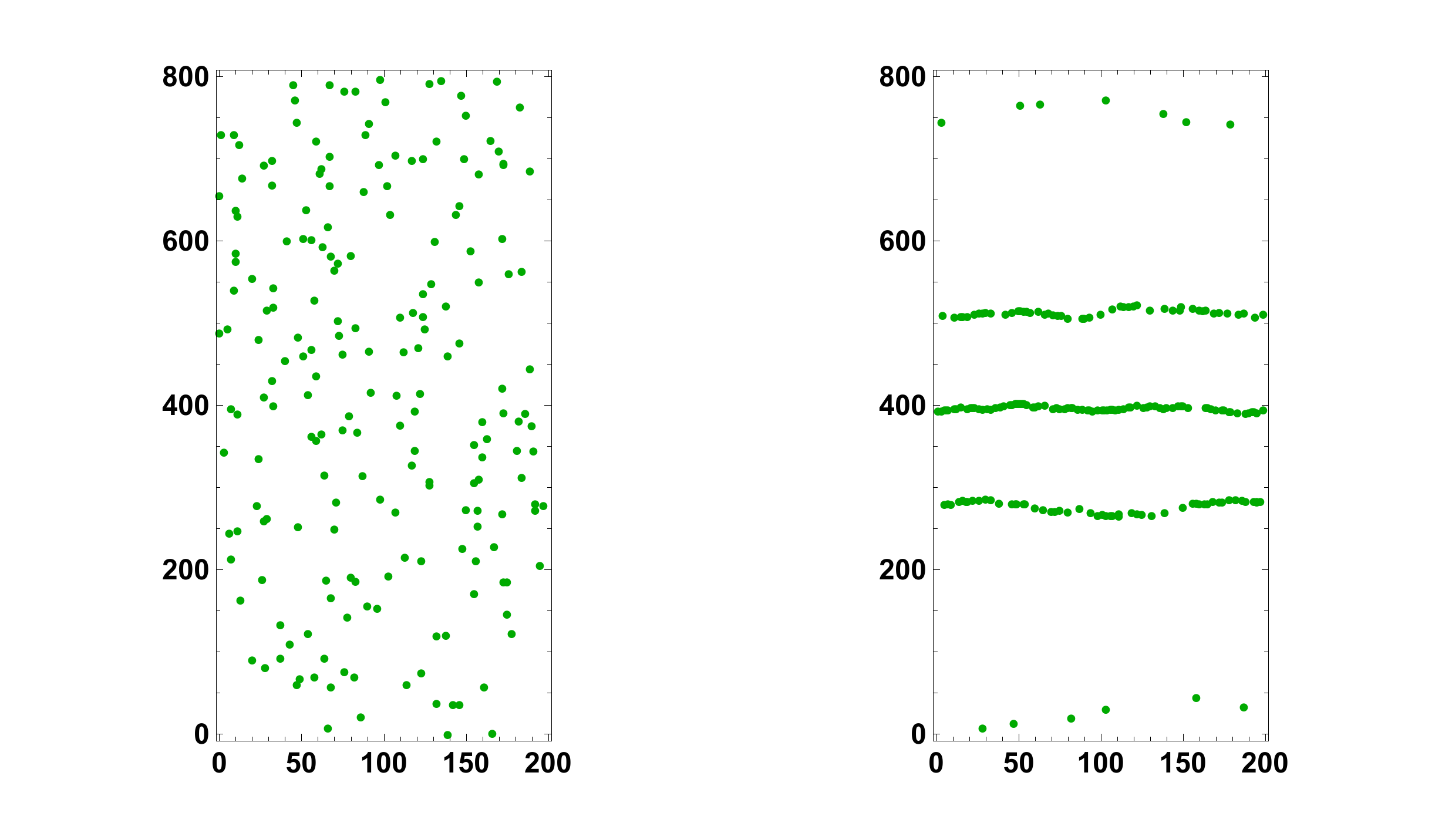}
  \caption{ An example of a $\dimm=2$ configuration for  $\beta=0.8$ (left) and $\beta=3.2$ (right).}\label{fig:2dsites} 
  \includegraphics[width=0.7\textwidth]{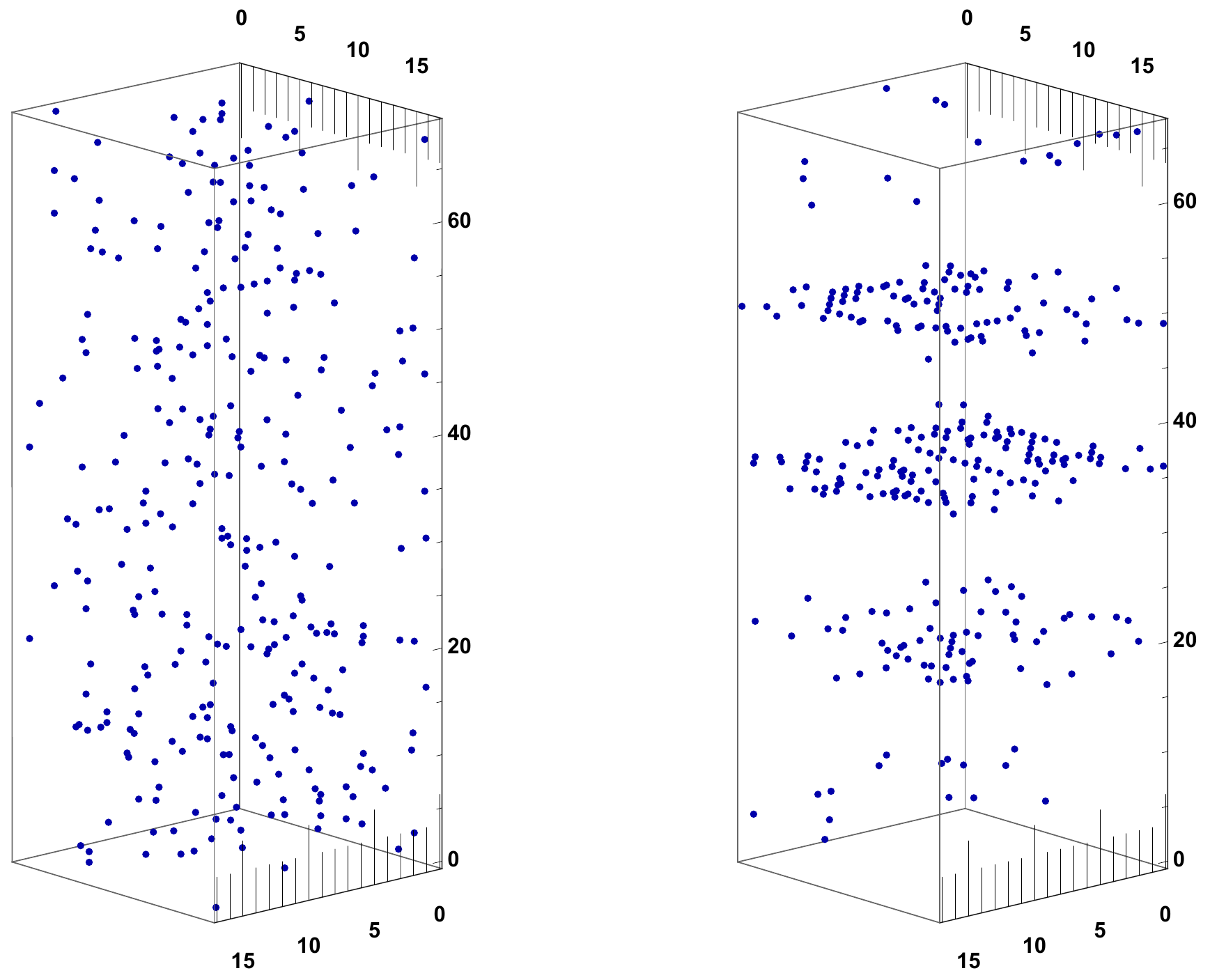}
  \caption{ An example of a $\dimm=3$ configuration for  $\beta=1.0$ (left) and $\beta=3.0$ (right).}\label{fig:3dsites} 
\end{figure}

Next, we process the raw data by first finding the autocorrelation times. In Fig.~\ref{fig:autocorr},  we show a typical
autocorrelation plot for the cold and hot values of $\beta$ in $\dimm=2,3$.  After finding the autocorrelation times from the raw
data, and discarding the first few autocorrelation times, we calculate the expectation value of the observables and the
standard error.  These are shown in Fig.~\ref{fig:observables}, making explicit our
surmise from the raw data, that the system undergoes a phase transition at some $\beta=\beta_c^{(\dimm)}$. For our
simulations, $\beta_c^{(2)}\approx 2.344$, and $\beta^{(3)}_c\approx 1.980$, consistent with our expectations from the raw
data.

As $\beta$ becomes much larger, one finds the expected critical slowing down, which means very large $\beta$ values
become more expensive to study. Moreover, around the transition values of
$\beta$, since thermalisation is poor, it is harder to extract reliable expectation values even for $10^4$
sweeps. The raw data shows that the sytem oscillates between the hot and cold states and hence,  rather than a Gaussian,
one has a double Gaussian. As $\beta$ increases one expects the larger peak to shift from the hot phase to the cold phase,
given a sufficient number of sweeps. However, the $10^4$  sweeps in our simulations are not always sufficient to see
this. Fig.~\ref{fig:badthermalisation} shows an example  of this in $\dimm=2$ for two values $\beta$ near
$\bctwo$. From our data, at the higher  value of $\beta=2.332$, the system spends all the $10^4$ sweeps in the
hot phase, while at the lower value  $\beta=2.328$ it oscillates between the hot and cold phases. This explains the
lack of consistency in the data in Fig.~\ref{fig:observables} around the phase transition.

It is also possible to calculate the specific heat $\mathcal{C}$ as a function of $\beta$, where  
\begin{equation}
\mathcal{C}= \beta^2\langle S-\langle S\rangle\rangle^2. 
\end{equation}
$\mathcal C$ can be seen to develop  a peak around the phase transition,  but again, because of poor data in this
region, we set aside this analysis for now. 
  

 In order to gauge the nature of a typical configuration, we show examples of  causal sets  that have been
 generated in the MCMC simulations in the hot and cold phases, in Figs.~\ref{fig:2dsites} and \ref{fig:3dsites}. In both  $\dimm=2,3$, the causal set in the hot phase is
 clearly   manifold-like in its random distribution, while in the cold phase it is layered and
 non-manifold-like. The number of layers in the cold phase in both $\dimm=2,3$  is approximately $5$, with most of the
 elements being distributed in the middle three layers. It is also clear from these figures that each element in a
 given layer is related to all the elements in the layer just above it and the one below it. 

\begin{figure}[t]
\centering
\includegraphics[width=\textwidth]{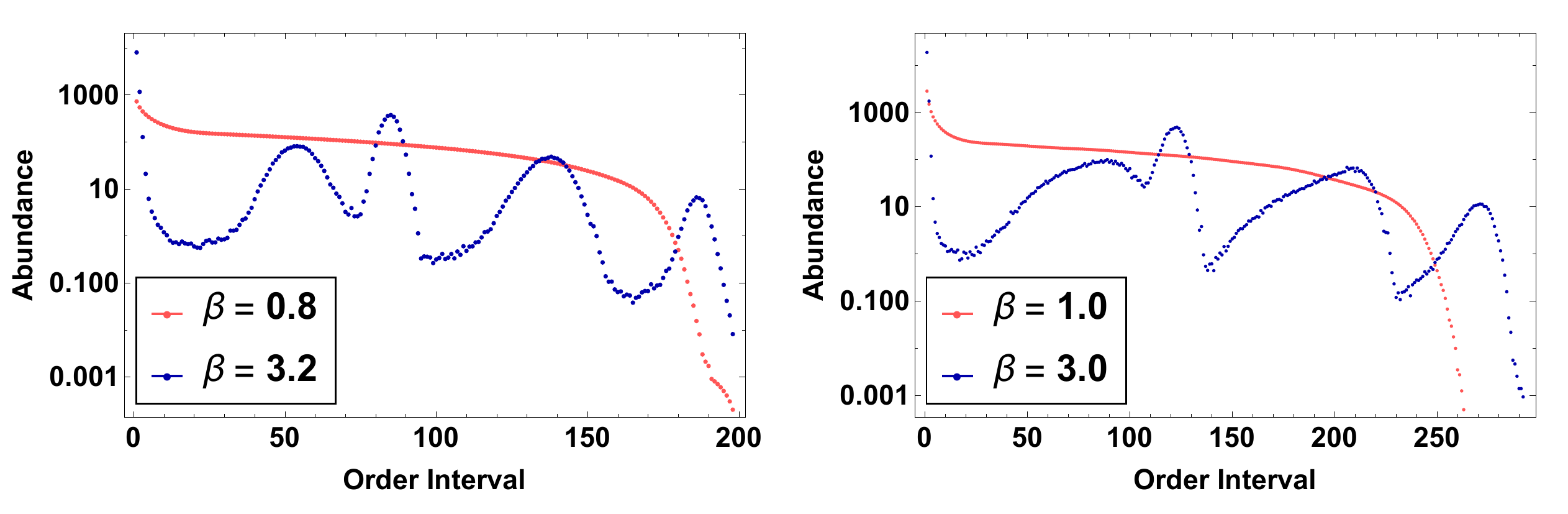}
\caption{The interval abundances for the two phases in $\dimm=2$ (left) and $\dimm=3$ (right).}
\label{fig:interval_abundances}
\end{figure}

 In Fig.~\ref{fig:interval_abundances},  we show the expectation values of the various interval abundances in the hot
 and cold  phases for  $\dimm=2$
  and $\dimm=3$.  This distribution is drastically different in the two phases. In the hot phase, it is a reflection of
  the  manifold-like properties of the causal set with, in particular,  the interval abundances decreasing
  monotonically with $k$, the size of the intervals \cite{intervals}.  In the cold phase, on the other hand, the number of
  links is very large and the distribution develops ``oscillations'', with 
  distinct peaks at very large values of $k$.   The abundance of
  intervals of size $k=54$ in $\dimm=2$ and $k=89$ in $\dimm=3$, for example correspond approximately to the size of
one of the layers. Such intervals lie between the layers $i-1$ and $i+1$. The next peak is sharper and higher and likely
corresponds to the intervals between $i=2$ and $i=4$, suggesting that the central layer $i=3$ is the densest.  

\begin{figure}[t]
\centering
\includegraphics[width=\textwidth]{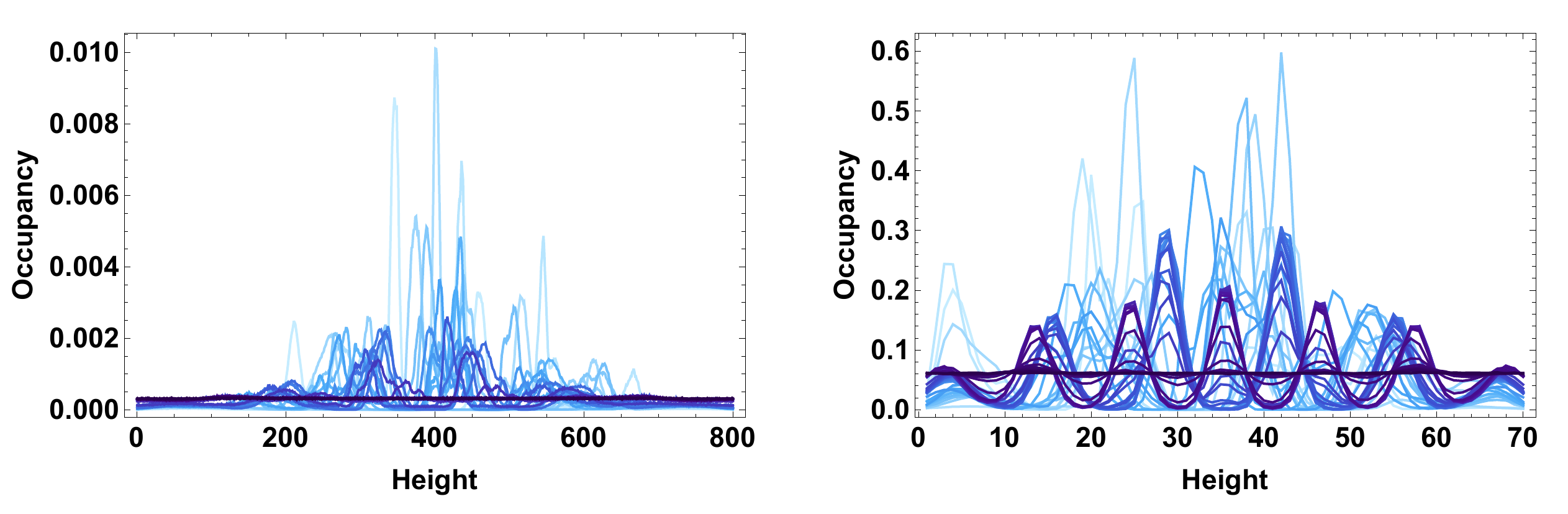}
\caption{Occupancy or the number of elements at each time slice, for different values of $\beta$ for $\dimm=2$, $n=200$ (left) and $\dimm=3$, $n=300$ (right). In both cases we take $\epsilon=0.1$ and average over $50$ causal sets each $200$ sweeps apart.}
\label{fig:occupancy}
\end{figure}

Fig.~\ref{fig:occupancy} illustrates the change in the occupancy of each constant time slice in the lattice, from a
random  manifold-like order to a layered order.  For  $\beta<\beta_c^{(\dimm)}$ the occupancy is uniform, whereas for
$\beta > \beta_c^{(\dimm)}$   it is concentrated on very few layers. As noted above, the  average number of  layers in the second phase for $\dimm=2$, $\beta=3.2$ is $\sim 5$, and we observe similar behavior for $\dimm=3$, $\beta=3.0$. The occupancy is not
uniform and tends to be largest in the middle layer, and smallest at the bottom-most and the top-most layer. 

The functions $f_{2,3}(k,\varepsilon)$  in Fig.~\ref{fig:fne} which appear  in the BD action,
Eqn.~(\ref{eq:bdactions},\ref{eq:fnes})  give a hint about the factors at play.  We  note that the
action contribution will be larger if there are only $k$-intervals  for which  $f_{2,3}(k,\varepsilon) > 0$. This is reflected by the abundance of intervals in Fig.~\ref{fig:interval_abundances}, though the appearance of tertiary and higher
peaks is a result of remnant entropic factors.  

The simulations slow down considerably for larger $\beta> \beta_c^{(\dimm)}$, and hence the data becomes
unreliable.  However, one finds a clear trend as $\beta$
increases in Fig.~\ref{fig:observables}, suggesting that the observables have not reached their asymptotic,  zero
temperature values. 

In the zero-temperature  phase  we expect the action to completely dominate 
the entropy, which, as argued in \cite{fss}, is the  symmetric maximally connected bilayer poset. Such posets have
approximately equal number of elements $\sim n/2$ in the two layers,  with every  element in the first
layer related to every element in the second layer, so that the number of links is $\sim n^2/4$ (and the ordering
fraction is $\sim 0.5$ for large $n$.).  The  symmetric maximally connected bilayer poset is the unique  $n$-element
poset in $\cEcylmn_\dimm$ which maximizes the number of links. The BD action in both $\dimm=2$ and $\dimm=3$   thus 
takes its smallest value for these  bilayer posets (Eqn.~(\ref{eq:bdactions},\ref{eq:fnes}), which hence dominates the
partition function.  The maximally connected bilayer poset can be thought of as the ``ground state'' or zero-temperature
state of this theory. 

\begin{figure}[t]
\centering
\includegraphics[width=\textwidth]{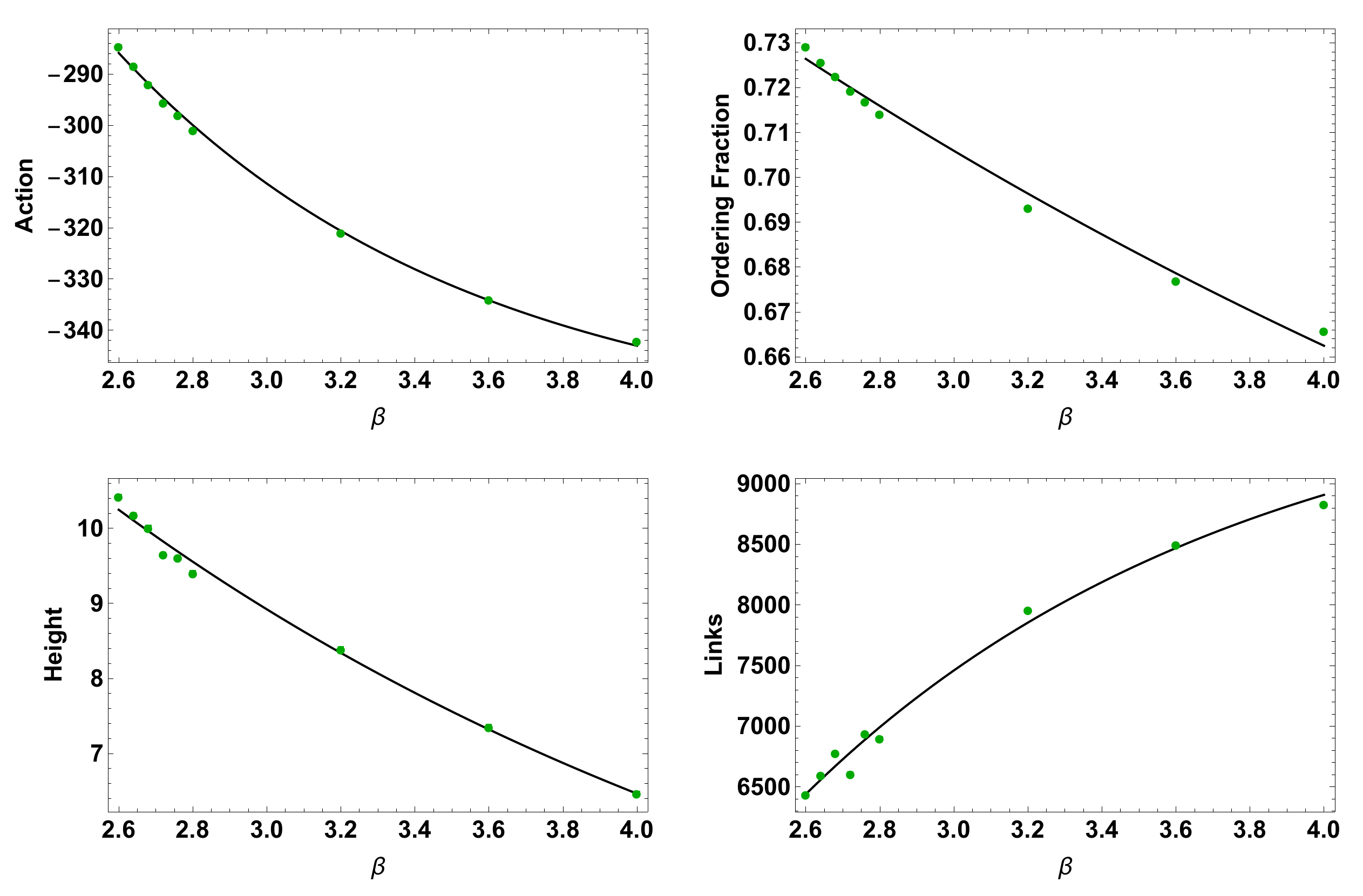}
\caption{The fitting functions take the form $f(\beta)=A+B\exp(-C\beta)$, where the asymptotic value $A$ is set to be
  the value of the corresponding observable for the  bilayer poset.}
\label{fig:asymptotics}
\end{figure}

We find that our data is consistent with this expectation. In Fig.~\ref{fig:asymptotics}, we use fitting functions in the
cold phase to model our data for each of the observables in $\dimm=2$.   For  $n=200$,  the number of links
in the symmetric  bilayer poset is   $10^4$,  while the BD action is $-360$. This is consistent with the  fitting
functions we find. For the $\dimm=3$ case  with $n=300$, the expectation value of the  number of links is $\sim 22,000$, and the BD
action is $\sim -380$. From Fig.~\ref{fig:observables},  we note that this is consistent with the data.

\section{Conclusions}
\label{sec:conclusions} 

The results presented in this work are a first step towards a full dimensional restriction of causal
set theory,  and  are the first explicit explorations of  non-trivial global spatial topology in $\dimm=2,3$.

For the theory defined by the partition function $\cZ_{n,m}^{(\dimm)}(\beta)$,  Eqn.~\ref{eq:partn_stat}, we see that as $\beta$
varies, there is a phase transition from a manifold-like phase to a non-manifold-like phase, which represents the battle
between entropy and the action,  and is similar  to that observed for the  $2$-orders \cite{2dqg,fss}.  The BD action in both
$\dimm=2,3$ is such that the measure is maximised by increasing the total number of links, which means that the ground
state (or $\beta \rightarrow \infty$) is the symmetric maximally connected bilayer poset.  For smaller values of $\beta$ where the
entropy-action battle is more subtle, however, there are 
interesting features which emerge.  
 
The most obvious question that arises in this work is the role played by the choice of continuum spacetime $(M,g)$ in
setting up the partition function $\cZ_{n,m}^{(\dimm)}(\beta)$. It is clear that the continuum
phase is dominated by this geometry, whereas the layered phase  seems largely independent of it. 
Importantly, the  choice of $(M,g)$  introduces a preferred relative measure  between manifold-like
causal sets in $\ocylmn_\dimm$, since those that are approximated by $(M,g)$ are far more numerous than those that are
not. Thus, at best, $\cZ_{n,m}^{(\dimm)}(\beta)$  represents fluctuations about the causal sets $c\sim (M,g)$, given that such a
restriction is meaningful.

We have demonstrated that the choice of lattice size $m$ is not relevant as long as $m$ is large enough. However we have
chosen a specific,  regular lattice,  and we don't yet know whether our results will change with a different lattice
geometry. A more natural choice for the background  lattice is a random $m$ element causal set obtained via a sprinkling
into $(M,g)$. It would be important to explore simulations on such lattices and compare with our current results. 

In order to obtain a more complete dimensional restriction,  one would need to vary
over all geometries with a given global topology and  subsequently, vary over all topologies. 
By allowing  the lattice $\mlcylm$ to vary,  the relative weights should resemble that expected from
the continuum. Work in this direction is currently underway \cite{varyinggeom}. Varying over topologies is trickier but
preliminary investigations in this direction have also begun.  

We have also restricted our analysis here to fixed values of $n$, without addressing the question of how robust the
results are in the asymptotic or large $n$ limit. As shown in \cite{fss} using a finite sized scaling analysis for the $\twod$
orders, one must look for scaling behaviour with $\beta$, $n$ and $\epsilon$. Work is currently being done towards this
goal \cite{2d3dfss}.

Finally, as in all related work \cite{2dqg,hh,fss}, there  is the question of the analytic continuation and the physical
significance of $\beta$ in full Lorentzian quantum gravity.  This remains an important open question. Clearly the
entropy-action struggle plays out differently for the quantum path integral as shown in \cite{carliploomis}. There, a
range of parameters was found for which the bilayer poset is suppressed. Relating such quantum results to the current
statistical ones would be useful while exploring different analytic continuations.  The hope is that
further explorations of the sample space of causal sets and different analytic continuations will help us eventually
arrive at a deeper understanding of these questions.    

\section*{Acknowledgements}
Research at Perimeter Institute is supported in part by the Government of Canada through the Department of Innovation,
Science, and Economic Development Canada and by the Province of Ontario through the Ministry of Economic Development,
Job Creation, and Trade. S. Surya is supported in part by a Visiting Fellowship at the Perimeter Institute. This
research was enabled in part by support provided by the Discovery cluster at Northeastern University
(\url{www.northeastern.edu/rc}), the MP2B cluster operated by Calcul Qu\'{e}bec (\url{www.calculquebec.ca}), the Cedar
cluster operated by WestGrid (\url{www.westgrid.ca}), Compute Canada (\url{www.computecanada.ca}), and Raman Research
Institute (\url{www.rri.res.in}). We would like to thank Lisa Glaser for discussions during the later part of this
work. 
\clearpage
\appendix
\section{Some Definitions for CST}
\label{app:defs} 
This section contains the definitions of various standard terms in CST that have appeared in the preceding sections. 

\begin{itemize}

 \item  A causal set $c$ is
  said to \emph{causally embed}   into a spacetime $(M,g)$  if there exists an injection $\varphi: c \hookrightarrow (M,g)$ such that  $e_1 \prec e_2 \leftrightarrow \varphi(e_1) \prec
  \varphi(e_2)$ for all $e_1,e_2 \in c$.  If the causal embedding is further obtained via a Poisson-sprinkling, then $c$ is
  said to be \emph{faithfully embedded}  into $(M,g)$, i.e.,  $c$ is said to be \emph{approximated} by the continuum spacetime,
  (denoted $c \sim (M,g)$) and is hence \emph{manifold-like}. Importantly, not all causal sets that causally embed into $(M,g)$
  are manifold-like.

\item A \emph{sample space} is a collection of causal sets. In this work we refer to the following sample spaces:
  $\Omega_n$, the set of all $n$-element causal sets, $\Omega_n(M,g)$ the set of all $n$-element causal sets that
  causally embed into $(M,g)$ and    $\Omega_\dimm^{(m,n)}$ the set of all $n$ element causal sets that embed into the lattice
  $\ocylmn_\dimm$.

\item An \emph{order invariant}  for a causal set is a
function on $c$ (over any field) which is independent of relabellings of $c$. In all the examples in this work, the
functions considered are only over the field $\re$. The \emph{order invariance} of a quantity (for example the measure) 
therefore means that it is independent of a relabelling of the causal sets.

 \item  An $n$ element  \emph{chain} is a completely ordered $n$-element set $c$, i.e., for every $e_i, e_j \in c$, either $e_i \prec e_j$ or
   $e_j \prec e_i$.   An $n$-element  \emph{antichain} is a set of mutually unrelated elements: $e_i\nprec e_j \,\,\forall \,\,e_i, e_j\in c$. 

 \item A \emph{layer} in a causal set $c$ is an antichain of elements with equal proper time, $\tau(e_i)$, defined as
   the longest chain from $e_i$ to any minimal element in the causal past of $e_i$.\footnote{ In the  low temperature
     phase of the partition function $\cZ_{(n,m)}^{(\dimm)}(\beta)$ where the action dominates over the entropy, we observe
     that causal sets self-assemble into layers in which nearly all elements are related to all those in adjacent layers (Figs.~\ref{fig:2dsites},~\ref{fig:3dsites}).}
   
   \item A   \emph{crown poset} $c$  is an  $n=2m, m \geq 3$ element poset with exactly two non-intersecting equal sized antichains
     $E_{1,2}$,  such that $c=E_1 \cup E_2$. Moreover, $c$ admits a labelling such that $\forall \, \,
     i\mathrm{(mod)}m, e_i^1\prec e_i^2 , e_{i+1}^2$, are  the only relations,  with $e_i^1\in E_1, e_{i}^2, e_{i+1}^2 \in E_2$.   A crown poset causally embeds into any spacetime with a spatial  non-contractible
     $S^1$ factor.
     \item The \emph{interval} $I[e_i,e_j]$ in a causal set $c \ni e_i, e_j$ is the set of elements $e_k \in c$ such
       that $e_i\prec e_k \prec e_j$.   Thus a \emph{$r$-element interval} $I[e_i,e_j]$ contains $r$ ``intervening''
       elements. For $e_i \prec e_j$, if  $I[e_i,e_j]=0$, $e_i,e_j \in c$ are said to be  \emph{linked}.  The \emph{abundance
         of $r$-element intervals}  $n_r$ in a causal set  $c$ is simply the total number of $r$-element intervals.

\item The \emph{ordering fraction} in an $n$-element causal set is the ratio of the number of relations $r$ to the total
  number of possible relations on $n$-elements which is $\binom{n}{2}$.

\item The \emph{height} of a causal set is the length of the longest chain in the causal set. 

\item The \emph{Benincasa-Dowker action} is an order invariant which converges to the Einstein-Hilbert action for causally convex compact regions of spacetime. In two and three dimensions, it is defined as
\begin{equation}
\begin{split}
S_{BD}^{(2)}(c)/\hbar&=2\epsilon\left[n-2\epsilon\sum\limits_{r=1}^{n-1}n_rf_2(r-1,\epsilon)\right]\,, \\
  S_{BD}^{(3)}(c)/\hbar&=\frac{1}{\Gamma(5/3)}\left(\frac{\pi\epsilon}{3\sqrt{2}}\right)^{2/3}\left[n-\epsilon\sum\limits_{r=1}^{n-1}n_rf_3(r-1,\epsilon)\right]\,,
                     \label{eq:bdactions} 
\end{split}
\end{equation}
where $f_{2,3}(r, \epsilon)$ are the smearing functions 
\begin{equation}
\begin{split}
f_2(r,\epsilon) &=
                     (1-\epsilon)^r\left[1-\frac{2r\epsilon}{1-\epsilon}+\frac{r(r-1)\epsilon^2}{2(1-\epsilon)^2}\right]\,,\\
f_3(r,\epsilon) &=
                     (1-\epsilon)^r\left[1-\frac{27r\epsilon}{8(1-\epsilon)}+\frac{9r(r-1)\epsilon^2}{8(1-\epsilon)^2}\right]\,.
                     \label{eq:fnes} 
\end{split}
\end{equation}
and $\epsilon=(0,1]$ is a non-locality scale, which is a free parameter in the theory.  
\begin{figure}[t]
  \centering
  \includegraphics[width=0.5\textwidth]{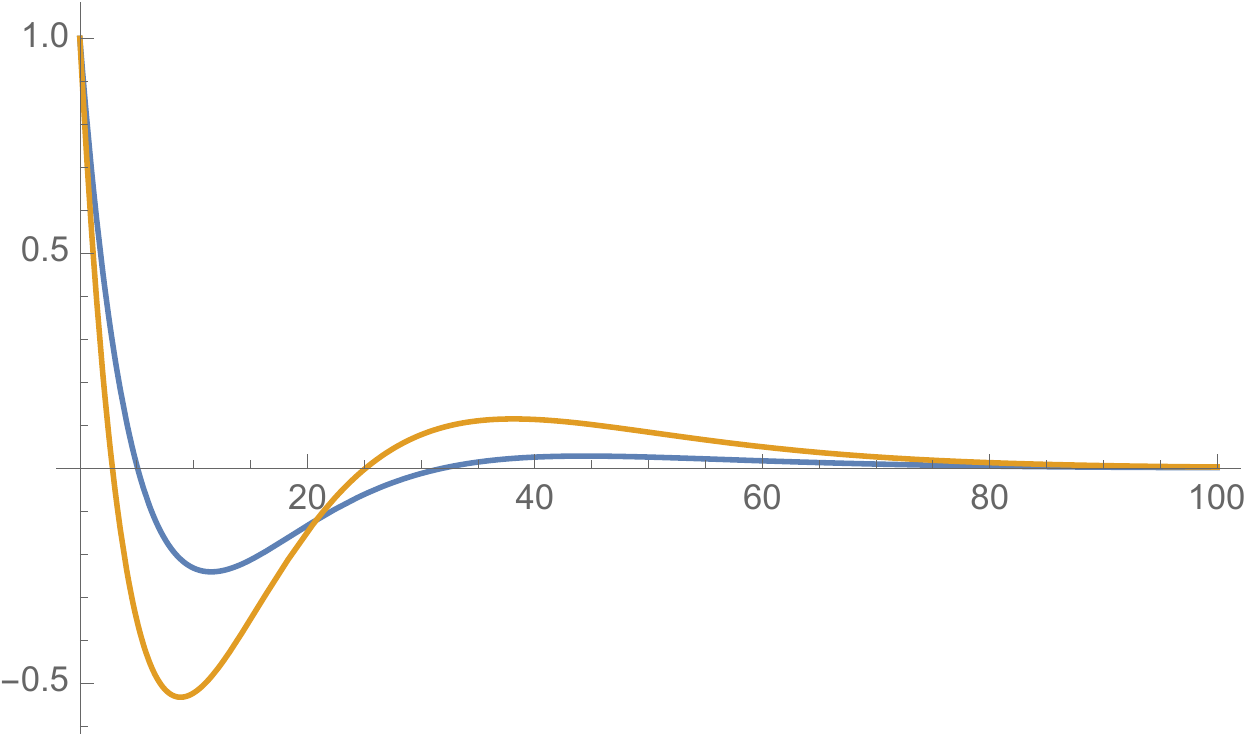}
  \caption{The functions $f_2(m,\epsilon)$ and  $f_3(m,\epsilon)$ for $\epsilon=0.1$ as a function of the
    interval size. There are three
    consecutive bands of interval sizes,  with alternating signs.}\label{fig:fne} 
\end{figure}

\item 
An $n$-element linear order $L=( l_1, \ldots l_n)$ over the set of
  $n$-integers $S=\{ 1, \ldots n\} $  is an automorphism of $S$ and is therefore totally ordered.  An  $n$-element poset
  $C$ is said to be a \emph{$\dor$-order} if it is the ``intersection''  of $n$-element linear orders
  $\{ L^1, \ldots, L^{\dor}\}$: i.e.,  if its elements are given by $ e_j \equiv (l_j^{1}, l_j^{2}, \ldots l_j^{\dor})$, such that  $e_j \prec  e_k$ iff
  $l^i_j<l^i_k$ for all $i$. $C$ is then said to be of order-theoretic dimension $\dor$  and has a natural
  representation in the $\dor$-dimensional hypercube.  
  The \emph{$2$-orders} (or $\twod$-orders)  in particular, admit a simple representation
  with $L^{1,2}$  being the light-cone coordinates $\{(l^1_i=u_i,l^2_i=v_i)\}$ of the elements $e_i=(u_i,v_i) $ in  a
  regular light-cone lattice in  $(D_2,\eta)$.

\end{itemize}









\section{\texorpdfstring{A calculation of the ordering fraction in $\dimm=2$}{A calculation of the ordering fraction in d=2}}  
\label{app:of} 

The ordering fraction in an $n$-element causal set is the ratio of the number of relations  $R$ to the total number of
possible relations $n(n-1)/2$. For  large $n$, $r \sim 2R/n^2$. We now calculate the expectation value of $R$ for causal
sets obtained via a Poisson sprinkling into the $\dimm=2$ cylinder spacetime $(M,g)$.

\begin{figure}[!t]
\centering
\includegraphics[width=0.5\textwidth]{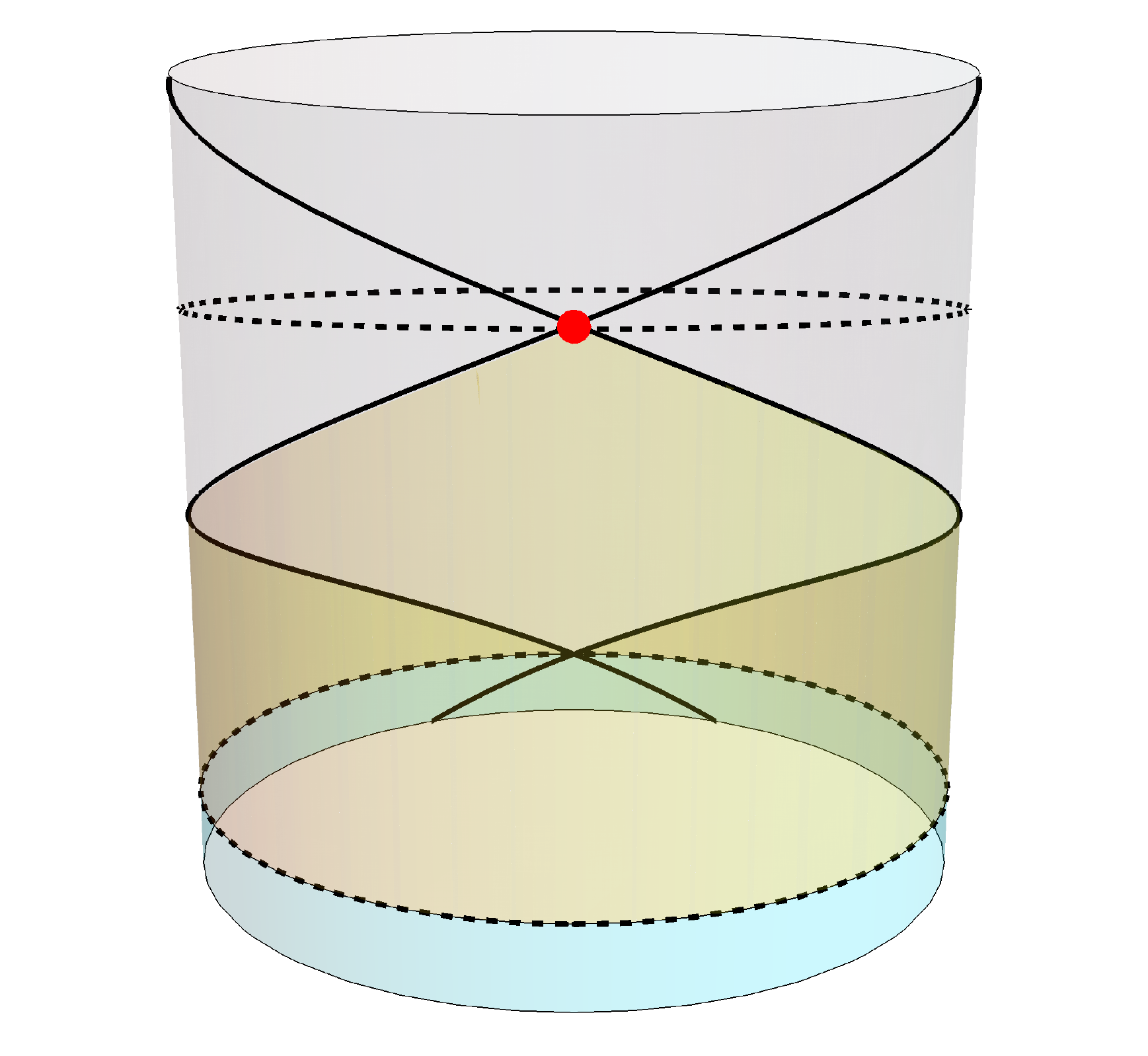}
\caption{When the aspect ratio $\alpha>1/2$, the light-cone for an element  can wrap all the way around the cylinder.}
\label{fig:cylinder_wrapping}
\end{figure}

Consider the finite region $M$, where $M\sim S^1 \times [0,h]$ for some choice of $h,w$, with
$ds^2=-dt^2+w^2d
\theta^2$, and the aspect ratio, $\alpha=h/w$.    Let $\phi: c \hookrightarrow
M$. Consider the event  $p=(t,x)=\phi(e)$, $e \in c$. The average  number of  events in $\phi(c)$ that are related to
$p$ is given by the volume of the region $I^+(p) \cap M$ and $I^-(p) \cap M$. The average number
of relations in the ensemble of causal sets that are approximated by $M$ is thus given by 
\begin{equation}
\av{R} = \int_{M} dp \int_{q\prec p} dq = \int_{M} dp \, \vol(I^-(p) \cap M)
\end{equation}
where we need to consider only the past of $p$,  $I^-(p) \cap M$, and we take $\rho=1$. 
While the ordering fraction for  $\mink^d$ is related to the continuum dimension via the Myrheim-Meyer
dimension estimator, and is hence independent of the size of the region,  this is not the case here. 
For $\alpha<1/2$, the light rays from $p$ do not wrap around the cylinder, and the result is similar to
$\mink^2$. For  $\alpha>1/2$ this is not the case and we must split the past volume  from $p$ into two different parts
as shown in Fig.~\ref{fig:cylinder_wrapping}.

For $p=(t,x)$, with $t \leq  w/2$, $V(p)\equiv \vol(I^-(p) \cap M)=t^2$. The average number of such relations is therefore given by
\begin{equation}
\int_{t=0}^{w/2} dt \int_{0}^w dx\, t^2 = w^4/24\,. 
  \end{equation} 
For $p=(t,x)$, $t\geq w/2$, the average number of relations splits up into two parts, one with (a) $q=(t',x')$ such that $t'
\leq  t-w/2$ and one with (b) $t' \geq t-w/2$ (see Fig.~\ref{fig:cylinder_wrapping}). Thus the average  number of relations for such $(p,q)$
is 
\begin{equation}
\int_{w/2}^h dt \int_0^w dx (V_a(t) +V_b(t))=  \int_{w/2}^h dt \int_0^w dx (w(t-w/2) +w^2/4)= w^2h^2/2-w^3h/4\,. 
  \end{equation} 
Adding these two expressions gives us the total expectation value for the number of relations 
\begin{equation}
  \av{R} = \frac{(hw)^2}{2} \biggl( 1-\frac{1}{2\alpha} + \frac{1}{12 \alpha^2} \biggr)  \,,
  \end{equation} 
  or
  \begin{equation}
 \av{r} = \biggl( 1-\frac{1}{2\alpha} + \frac{1}{12 \alpha^2} \biggr). \,.
    \end{equation} 

In our simulations in $\dimm=2$, we have used $\alpha=4$. This corresponds to an $\av{r}\sim 0.88$, which is consistent
with our observations.

\bibliography{Masterref}
\bibliographystyle{apsrev4-1}
\end{document}